%% file: paper.tex
\documentclass[submission, Phys]{SciPost}

\pdfoutput=1
\input{incl_settings.tex}
\input{incl_shortcuts.tex}

\graphicspath{{./figs/}}

\begin{document}

    \begin{center}
        {\Large \textbf{\color{scipostdeepblue}{
        Unfolding Weak Boson Fusion
        }}}
    \end{center}

    \begin{center}
      Javier Mariño Villadamigo\textsuperscript{1},
      Lydia Brenner\textsuperscript{2},
      Anja Butter\textsuperscript{1,3}, 
      Lisa Marie Lehmann\textsuperscript{2}, \\
      Bogdan Malaescu\textsuperscript{3}, 
      Sofia Palacios Schweitzer\textsuperscript{4}, and
      Tilman Plehn\textsuperscript{1,5}
    \end{center}

    \begin{center}
      {\bf 1} Institut für Theoretische Physik, Universität Heidelberg, Germany \\
      {\bf 2} Nikhef - Dutch National Institute for Subatomic Physics, Amsterdam, The Netherlands\\
      {\bf 3} LPNHE, Sorbonne Universit\'e, Universit\'e Paris Cit\'e, CNRS/IN2P3, Paris, France \\
      {\bf 4} NHETC, Department of Physics \& Astronomy, Rutgers University, Piscataway, NJ, USA \\
      {\bf 5} Interdisciplinary Center for Scientific Computing (IWR), Universit\"at Heidelberg, Germany \\
    \end{center}

    \begin{center}
        \today
    \end{center}
    
    \section*{Abstract}
    {Generative unfolding turns reconstructed events into high-dimensional, unbinned particle- or parton-level event distributions. We apply it to WBF Higgs production, with its characteristic tagging jets and a central electroweak process. The distinctive tagging-jet kinematics motivates unfolding jets to hard partons. We unfold the full event kinematics, including decay neutrinos, with a dedicated phase space parameterization. A model-conditional generative network quantifies and controls the bias when unfolding events with dimension-6 Higgs couplings. Given the low signal-to-background ratio, we combine unfolding with a classifier-based background reweighting. This lets us unfold WBF Higgs events precisely and competitively infer a Wilson coefficient from unfolded signal distributions.
    }


    \vspace{4pt}
    \noindent
    \rule{\textwidth}{1pt}
    \tableofcontents
    \thispagestyle{fancy}
    \noindent
    \rule{\textwidth}{1pt}
    \vspace{10pt}

\section{Introduction}

The precision program at the LHC is one of the great achievements of experimental and theoretical particle physics, as it establishes the fundamental theory of elementary particles, including quantum corrections. Precision first-principle simulations, from the Lagrangian to detector-level events, are crucial for this program~\cite{Heinrich:2020ybq,Campbell:2022qmc} and are becoming a critical bottleneck~\cite{Butter:2022rso}.

Data unfolding solves this problem by defining appropriate data representations for a given analysis and reduces the computational effort from a large number of end-to-end forward simulations to a single inference~\cite{Cowan:2002in}. This $(i)$ enables efficient tests of multiple theory hypotheses; $(ii)$ makes it easy to combine analyses from different experiments; $(iii)$ allows experiments to publish their data in a meaningful way; $(iv)$ expands the LHC analysis community to capable theorists; and $(v)$ allows the theory community to implement improved predictions without a full data re-analysis.

Traditional unfolding methods~\cite{Lucy:1974yx,DAgostini:1994fjx,Hocker:1995kb,Schmitt:2012kp} rely on binning and typically fail for more than three phase space dimensions. Modern machine learning (ML)~\cite{Plehn:2022ftl} enables unbinned
unfolding in many dimensions~\cite{Huetsch:2024quz}. Two different approaches have emerged: discriminative methods, like
OmniFold~\cite{Andreassen:2019cjw} or its AUSSIE improvement~\cite{Ore:2026qgp}, train classifiers to reweight a reference simulation to the unfolded distribution~\cite{Chan:2023tbf,Desai:2024yft,Zhu:2024drd,Falcao:2025jom}. OmniFold has been applied to several experimental
analyses~\cite{LHCb:2022rky,H1:2023fzk,ATLAS:2024rpl,ATLAS:2025qtv,CMS:2025sws,Electron-PositronAlliance:2025hze}; alternatively, conditional generative networks can be trained to sample unfolded
events~\cite{Bellagente:2019uyp,Bellagente:2020piv,Vandegar:2020yvw,Howard:2021pos,Backes:2022sph,Shmakov:2023kjj,Diefenbacher:2023wec,Shmakov:2024gkd,Pazos:2024nfe,Butter:2024vbx,Favaro:2025psi,Butter:2025via,Butter:2025mek,Petitjean:2025tgk}. Both methods have complementary advantages.

To describe the properties of the fundamental particles and their interaction in and beyond the Standard Model, we employ the effective field theory extension of the Standard Model (SMEFT)~\cite{Buchmuller:1985jz,Leung:1984ni,Degrande:2012wf,Brivio:2017vri}, with a set of operators that allows, for instance, CP violation. Weak boson fusion (WBF) Higgs production comes with the major theoretical advantage of an electroweak process~\cite{Barger:1994zq,Rainwater:1996ud}, reflected in the particularly precise and reliable theory predictions~\cite{Arnold:2008rz,Ciccolini:2007ec,Bolzoni:2010xr,Frixione:2013mta,Cacciari:2015jma,Dreyer:2016oyx,Cruz-Martinez:2018rod,Asteriadis:2021gpd,Asteriadis:2023nyl,vanBeekveld:2023chs,Becher:2026uih}. In the SMEFT it combines two handles~\cite{Dawson:2018dcd}: the tagging jets probe the Lorentz structure of the Higgs coupling to weak bosons~\cite{Plehn:2001nj,Hankele:2006ma,Englert:2012xt}, whereas the Higgs decay gives us precise access to many different Higgs couplings with the same, fully correlated production process~\cite{Lafaye:2009vr}. 

Its unique physics benefits make WBF Higgs production an exciting unfolding opportunity. Unfolding jets is a conceptual challenge because their unfolding target is not well defined in perturbative QCD. The reason why it might work for WBF is the suppression of central QCD jet radiation. The Higgs decay to leptonically decaying $W$-bosons~\cite{Rainwater:1999sd,Kauer:2000hi} requires us to unfold the missing neutrino momentum components. With this decay, the top pair and continuum $WWjj$ backgrounds dominate over the signal even after a dedicated selection~\cite{Kauer:2000hi,ATLAS:2023pwa}. Technically, we use conditional flow matching (CFM)~\cite{lipman2023flow}, combined with a transformer, to encode all correlations~\cite{Butter:2023fov,Huetsch:2024quz}. A critical question is how well we can control the bias from the Higgs signal model on the unfolding networks, a challenge of a successful unfolding program.

We begin with a description of our signal and background processes and the simulated training data in Sec.~\ref{sec:sim}. In Sec.~\ref{sec:wbf_sig} we construct an appropriate phase-space parameterization to precisely unfold lepton and jet observables. In Sec.~\ref{sec:wbf_eft} we unfold data with an assumed dimension-6 contribution, first with networks trained on different simulations and then with a network trained conditionally on a range of predictions. The latter gives us a new way to quantify the corresponding systematics in Sec.~\ref{sec:wbf_env}. In Sec.~\ref{sec:back}, we show how backgrounds can be removed using a trained classifier, allowing us to unfold an assumed signal sample. Finally, in Sec.~\ref{sec:para} we compare the measurement of the CP-odd Wilson coefficient from a one-dimensional distribution at the reco level and unfolded to the parton level. 

We emphasize that our study builds on Ref.~\cite{Butter:2025mek}, so we skip all aspects covered in this earlier study and focus on the estimation of the simulation bias and on the background removal.

\section{Event generation and selection}
\label{sec:sim}

\begin{table}[b!]
\centering
\resizebox{\textwidth}{!}{
\begin{tabular}{lc|lc}
\toprule
pre-selection &   
& fiducial region & \\
\midrule
\# leptons & 2   &
\# $b$-jets ($p_T > 20~\gev, |\eta_{j}|<2.5$) &0  \\

(sub-)leading lepton & $p_T > 22(15) \gev, |\eta|<2.5$ &
\# jets ($p_{T,j} > 30~\gev, |\eta_{j}|<4.5$) & $\geq 2$  \\

lepton pair flavors & $e / \mu $   &
central jet veto ($p_T>20$\gev) & \checkmark   \\

lepton pair charge & 0    &
outside lepton veto & \checkmark  \\

lepton separation & $\Delta R(\ell,\ell) > 0.1$   &
jet inv mass & $m_{jj}> 450 \gev$  \\

lepton inv mass & $m_{\ell \ell} > 10 \gev$   &
jet separation & $|\Delta \eta_{jj}| > 2.1$  \\

\# jets & $\geq 2$  &
jet separation & $|\Delta \phi_{\ell \ell}| < 1.4$~rad  \\

jet momentum/eta & $p_{T,j} > 20~\gev, |\eta_{j}|<4.5$ \\

lepton-jet separation & $\Delta R(\ell,j) > 0.4$   \\
\bottomrule
\end{tabular}
}
\caption{Event selection following Ref.~\cite{ATLAS:2023pwa}. Both, the detector-level and the unfolding-target selection have to be fulfilled for an event pair to pass the cuts.}
\label{tab:selection}
\end{table}

We simulate all events with  MadGraph~\cite{Alwall:2014hca} and NNPDF parton densities~\cite{Ball:2012cx}, using Pythia~\cite{Bierlich:2022pfr,Skands:2014pea} for parton shower and hadronization, and Delphes~\cite{deFavereau:2013fsa} with the default ATLAS card as fast detector simulation.  We reconstruct jets using the anti-$k_T$ algorithm~\cite{Cacciari:2008gp} implemented in FastJet~\cite{Cacciari:2011ma}. All processes are simulated to leading order, as the technical challenge of our unfolding study is the variable and high multiplicity of the final state, not the moderate fixed-order corrections. 

Our signal process is weak boson fusion (WBF) on-shell Higgs boson production with a decay to two leptonically decaying $W$-bosons,
\begin{align}
pp 
&\to H \, jj + X \notag \\
&\to W^+_\mu W^-_e  \; jj + X 
\to (\mu^+ \nu_\mu) \;  (e^- \bar{\nu}_e)  \; jj + X \; .
\end{align} 
We include the charge-conjugated final state, and $X$ denotes additional jets radiated off the initial or final state. The two tagging jets have sizeable transverse momenta, large invariant mass, and a wide rapidity separation. To improve the simulation efficiency of the signal, we require $m_{qq} > 400$~GeV for the tagging quarks generated with the hard process. The total signal cross section is $0.26$~fb. 

To test the signal kinematics we include dimension-6 operators using the SMEFTsim model\cite{Brivio:2020onw}, including the SM interference and the dimension-6 squared terms. We focus on one CP-even and one CP-odd modification of the $HWW$ interaction,
\begin{align}
\ope_{HW} &= H^{\dagger}HW^i_{\mu\nu}W^{i\mu\nu} \notag \\
\ope_{\widetilde{HW}} &= H^{\dagger}H\widetilde{W}^i_{\mu\nu}W^{i\mu\nu} \; .
\end{align}

The leading background before cuts is top pair production with jets,
\begin{align}
pp 
&\to t \bar{t} + X \notag \\
&\to (b e^- \bar{\nu}_e) \; (\bar{b} \mu^+ \nu_\mu) + X  \; .
\end{align}
To improve the simulation efficiency we include a generation-level cut on the transverse lepton momenta. The LO cross section of 7.96~pb is many orders of magnitude larger than the signal. The second background, increasingly important after the tagging jet selection, is the continuum
\begin{align}
pp 
&\to W^+_\mu W^-_e \, jj + X \notag \\
&\to (\mu^+ \nu_\mu) \; (e^- \bar{\nu}_e) \, jj + X \; .
\end{align}
Again, we apply a generation-level cut on the transverse lepton momenta and the same $m_{jj}$ cut as for the signal. 

The event selection defining the signal region, Tab.~\ref{tab:selection}, follows the ATLAS analysis in Ref.~\cite{ATLAS:2023pwa}. Here, event selection refers to cuts applied to signal and backgrounds after MC generation, at the parton, particle, or detector level. We show the detailed cut flow in Tab.~\ref{tab:cutflow_parton}.

\begin{table}[t]
\centering
\resizebox{\textwidth}{!}{
\begin{tabular}{l@{}rrrrrrr}
\toprule
Cuts & \multicolumn{5}{c}{Higgs signals} & \multicolumn{2}{c}{backgrounds} \\
& SM & $\ope_{HW} (0.5)$ & $\ope_{HW}(-0.5)$ & $\ope_{\widetilde{HW}}(5.0)$ & $\ope_{\widetilde{HW}}(-2.0)$ & $t\bar{t}$ & $WWjj$ \\
\midrule
Initial \# events
& 100000 & 100000 & 100000 & 100000 & 100000 & 100000 & 100000 \\
$p_{T,j} > 20~\gev, |\eta_{j}|<4.5$
& 99691 & 99698 & 99728 & 99868 & 99733 & 99773 & 99641 \\
$p_{T,\ell} > 15~\gev, |\eta_\ell|<2.5$
& 77291 & 78094 & 77417 & 78596 & 78114 & 82549 & 90905 \\
$n_\ell = 2$
& 37408 & 38164 & 37181 & 43120 & 39374 & 33893 & 49666 \\
$n_e =1, n_{\mu} = 1$
& 36883 & 37645 & 36672 & 42659 & 38865 & 32321 & 49655 \\
$p_{T,\ell_1} > 22~\gev$
& 35838 & 36624 & 35617 & 42015 & 37954 & 32130 & 49460 \\
$\Delta R(\ell_1,\ell_2)>0.1$
& 35062 & 35704 & 34828 & 39125 & 36481 & 32021 & 49423 \\
$m_{\ell \ell}>10~\gev$
& 33181 & 33774 & 32944 & 37540 & 34719 & 31793 & 49413 \\
$\Delta R(\ell,j)>0.4$
& 33155 & 33749 & 32908 & 37522 & 34699 & 31741 & 49411 \\
$n_j \geq 2$
& 30536 & 31133 & 30381 & 36176 & 32551 & 29244 & 46349 \\
\midrule
$b$ veto
& 29534 & 30122 & 29358 & 34509 & 31316 & 4596 (29244) & 42857 \\
Central jet veto
& 21792 & 22203 & 21647 & 24587 & 22809 & 3587 (29244) & 28087 \\
Outside lepton veto
& 18200 & 18547 & 17879 & 17733 & 18092 & 841 (7364) & 8547 \\
$n_j \geq 2$
& 15948 & 16241 & 15722 & 16280 & 16106 & 429 (4179) & 7085 \\
$m_{jj}>450$\gev
& 13394 & 13620 & 13259 & 14375 & 13763 & 30 (511) & 4450 \\
$|\Delta \eta_{jj}|>2.1$
& 12972 & 13198 & 12784 & 13153 & 13070 & 11 (125) & 3709 \\
$|\Delta \phi_{\ell \ell}|<1.4$ rad
& 11362 & 11642 & 11147 & 12181 & 11675 & 3 (35) & 1444 \\
\bottomrule
\end{tabular}
}
\caption{Cut flows for a hypothetical 100k events each. For the SMEFT we assume $c_{HW}/\Lambda^2=0.5/\text{TeV}^2$, $c_{HW}/\Lambda^2=-0.5/\text{TeV}^2$ and $c_{\widetilde{HW}}/\Lambda^2=5.0/\text{TeV}^2$, $c_{\widetilde{HW}}/\Lambda^2=-2.0/\text{TeV}^2$. We apply all cuts at reco and parton levels and show the reco-level event counts. For $t\bar{t}$, the numbers in parentheses are without the $b$-jet and central jet vetoes, as used for the classifier training. The vetoes are then included as a global rescaling of the $t\bar{t}$ rate to match that of the vetoed sample.}
\label{tab:cutflow_parton}
\end{table}

\subsubsection*{Probabilistic vetoes}

\begin{figure}[t]
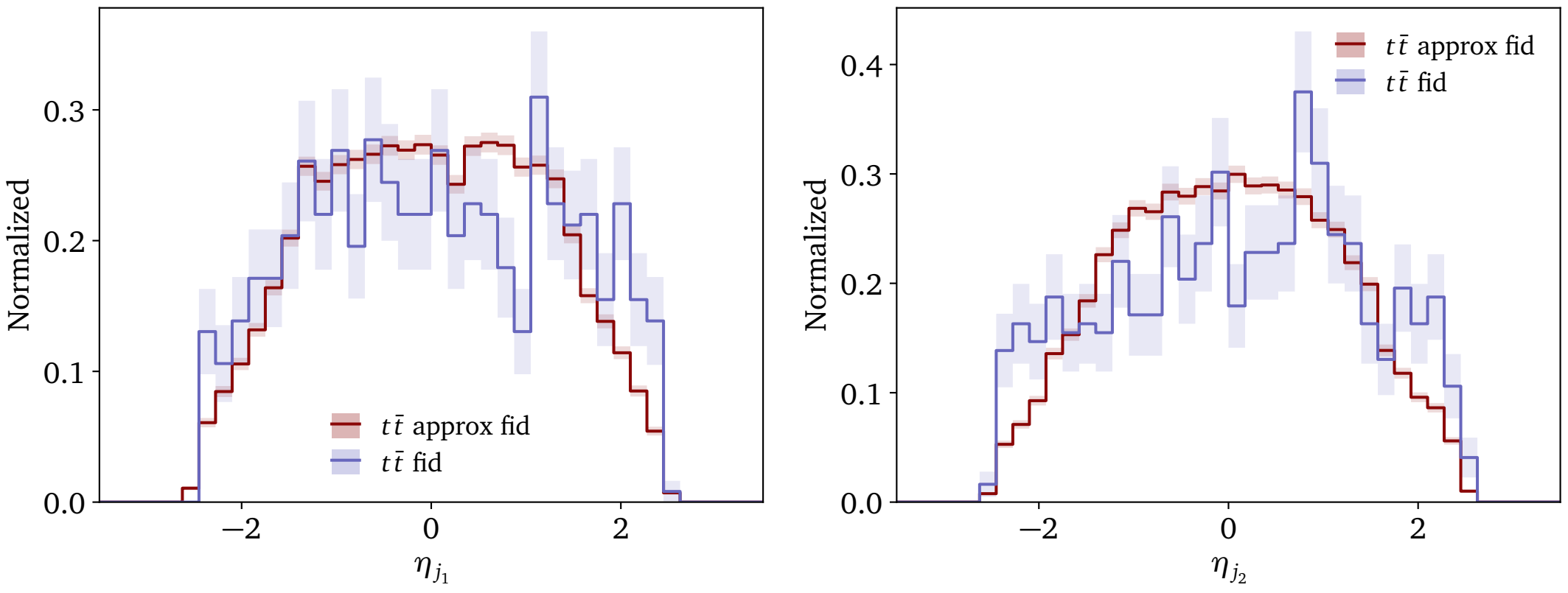

    \includegraphics[width=0.47\textwidth, page=11]{figs/ttbar_diff_hard_nobjet_nocjv.pdf}
    \includegraphics[width=0.47\textwidth, page=15]{figs/ttbar_diff_hard_nobjet_nocjv.pdf}
    \caption{Fiducial and approximately fiducial selections for $t\bar{t}$ background, where the approximate selection skips the central jet veto and the $b$-jet veto.}
    \label{fig:vetobackgrounds}
\end{figure}

According to Tab.~\ref{tab:cutflow_parton}, the selection rejects most of the background, in particular for top pair production. This improves the signal-to-background ratio, but leaves only few background events for training the background classifier in Sec.~\ref{sec:back}. A training sample comparable in size to the signal would require generating many more unweighted background events.

Two handles to solve this problem are the central jet veto, for which we do not consider events with additional jets between the two tagging jets in rapidity and the veto on $b$-tagged jets. For this study, we skip these two selection vetoes for the top background and replace them with a global probability. In Fig.~\ref{fig:vetobackgrounds} we illustrate its effect for the rapidity of the tagging jets in the top-pair background, one of the most sensitive observables for this selection. The kinematic shape of the fiducial and the $t\bar{t}$ events without these selection steps agree, while the uncut sample has considerably smaller statistical uncertainties. Using this approximation we find similar training statistics for our signal and background processes, around 2M Higgs signal events for each SMEFT parameter choice, 0.93M top pair events passing the particle-level selection, with an effective size of 0.15M after the reweighting to the parton-level selection described below, and 750k continuum background events.

For the $t\bar{t}$ background, the tagging jets typically do not originate from the top decay but from additional QCD radiation~\cite{Kauer:2000hi}. We cannot enforce the same generator cuts as for the signal and continuum background events. Instead, we build the parton-level event after generation, select the tagging quarks from a list of ISR partons and then apply the parton-level cut on them. However, the cuts are more stringent at parton level than they are at particle level, and even with the same reco-level selection the two samples have very different reco-level distributions. Therefore, generating a much larger sample inside the parton-level fiducial region is very inefficient.  Since the background events only enter our algorithm at reco level, we instead reweight all events passing the particle-level selection to those also passing the parton-level cuts at reco level, using an ensemble of five classifiers with weights normalized to a mean of one, as described in App.~\ref{sec:hyperparams}. The reweighted sample has an effective size of $17\%$, eight times more than the initial $2\%$ passing both selections.

\subsubsection*{Higgs-CP observable}

\begin{figure}[b!]
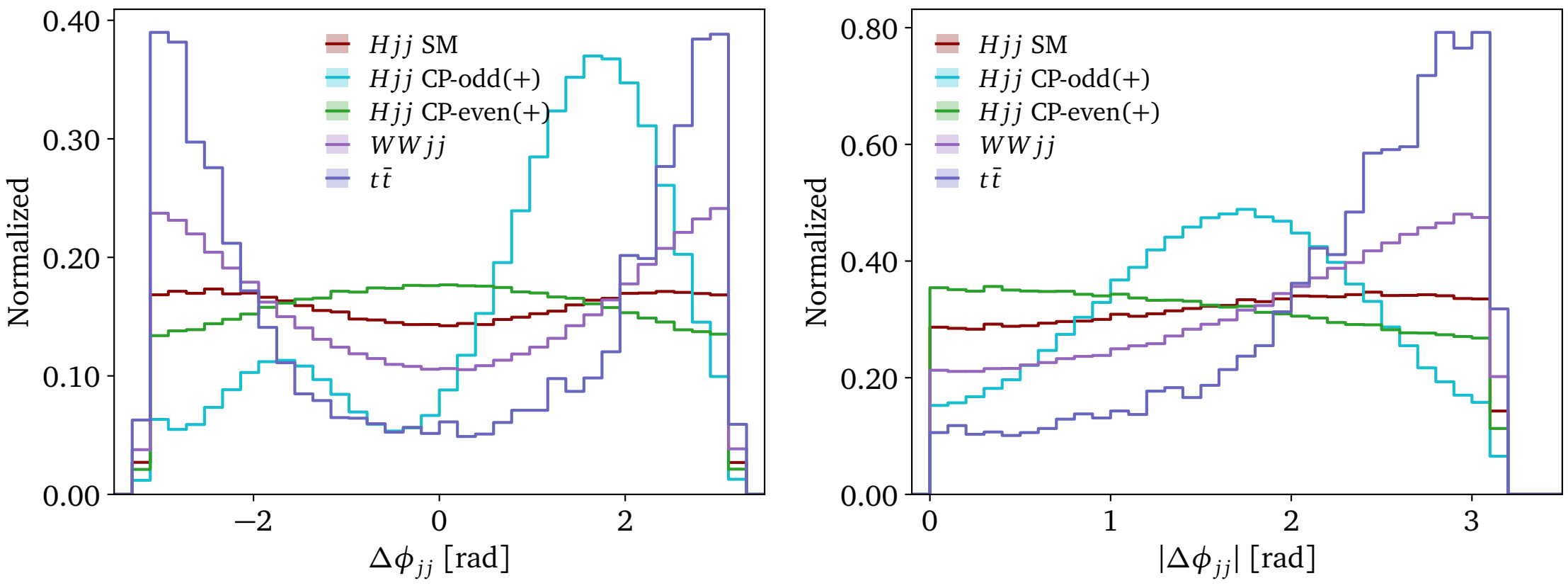

    \includegraphics[width=0.49\textwidth, page=11]{figs/dphijj_paper.pdf}
    \includegraphics[width=0.49\textwidth, page=13]{figs/dphijj_paper.pdf}
    \caption{Signed difference of the azimuthal angle between the two $\eta$-ordered tagging jets, the optimal observable for CP violation in WBF Higgs production.}
    \label{fig:dphijj}
\end{figure}

Finally, we illustrate the effect of the SMEFT operators inside the fiducial region. In Fig.~\ref{fig:dphijj} we show the signed azimuthal angle between the two tagging jets~\cite{Plehn:2001nj,Hankele:2006ma} as the optimal observable for CP violation in WBF Higgs production~\cite{Brehmer:2017lrt,Butter:2021rvz} at reco level. Although we see a clear difference in the normalized signal and background distributions, this difference will be weakened by the small signal-to-background ratio.

\section{Generative WBF Unfolding}
\label{sec:wbf}
We are interested in unfolding a reco-level measurement $\pd(\xr)$ to obtain a parton-level (pre-showering) or particle-level (post-hadronization) distribution by integrating 
\begin{align}
    p_{\text{unfold}}(\xp) = \int \d\xr\; p(\xp|\xr)\; \pd(\xr) \; .
\label{eq:def_unfold}
\end{align}
We focus on unfolding to parton level, but the following also holds for particle-level unfolding. A way to estimate this integral is to sample from a learned conditional probability, trained on forward-simulated event pairs $(\xp, \xr)$, 
\begin{align}
    \label{eq:conditional_density}
    \pmd(\xp|\xr) \simeq \psim(\xp|\xr)\; .
\end{align}
This requires detector universality, \ie the forward simulation $\psim(\xr|\xp)$ of the paired training events has to be independent of the physics process. Even then, the learned posterior inherits a prior dependence from the parton-level distribution $\psim(\xp)$ underlying the training data,
\begin{align}
    \pmd(\xp|\xr) \simeq \psim(\xp|\xr) = \frac{\psim(\xr|\xp)\; \psim(\xp)}{\int \d \xp'\; \psim(\xr|\xp')\; \psim(\xp')} \; .
    \label{eq:prior_dependence}
\end{align}
If the parton-level distribution of the measured data differs from $\psim(\xp)$, the unfolded distribution falls back to the training prior in all regions of phase space where the reco-level event does not fully constrain the parton-level kinematics. We study this simulation dependence in Sec.~\ref{sec:wbf_eft}.
As our generative network we use a conditional flow matching (CFM) network~\cite{lipman2023flow}. It transforms samples from a latent Gaussian distribution into samples over parton-level phase space, given a reco-level event. For this, it defines a time-dependent density that interpolates between the two distributions,
\begin{align}
 p(x,t|\xr) \to 
 \begin{cases}
   \psim(x \equiv \xp|\xr)  \qqquad & t \to 0 \\
   \normal(x \equiv r;0,1) \qqquad & t \to 1 \; . 
\end{cases} 
\label{eq:fm_limits}
\end{align}
Trajectories $x(t)$ are described, through an ordinary differential equation, by a velocity field
\begin{align}
 \frac{\d x(t)}{\d t} = v(x(t),t|\xr) \; ,
\end{align}
which is connected to the density via the continuity equation~\cite{Plehn:2022ftl}
\begin{align}
 \frac{\partial p(x,t|\xr)}{\partial t} + \nabla_x \cdot \left[ p(x,t|\xr)\, v(x,t|\xr) \right] = 0\; .
\end{align}
To encode the velocity field in a network, we exploit the fact that regressing against the conditional velocity $v(x,t|\xp, r)$, \ie with an additional conditioning on the endpoints of a single trajectory, produces the same expected gradient as regressing against the marginal velocity $v(x,t|\xr)$~\cite{lipman2023flow}. We construct this conditional velocity using a straight trajectory
\begin{align}
  x(t | \xp, r) = (1-t) \,\xp + t \, r 
  \qquad \Rightarrow \qquad 
  v(x(t),t|\xp, r) =  r - \xp \; .
\end{align}
We then learn the marginal velocity $v_\theta(x,t | \xr)$ with a simple MSE loss 
\begin{align}
    \loss_\text{CFM} = \Big\langle \left \|v_\theta(x(t), t|\xr) - (r - \xp) \right\|^2 \Big\rangle_{t \sim \mathcal{U}(0,1), (\xp, \xr) \sim \psim(\xp,\xr), r\sim \normal(0,1)} \;.
\end{align}
With this learned velocity, we use an adaptive ODE solver to generate unfolded  events for a given $\xr$ via
\begin{align}
    \xp = r - \int_0^1 \d t\, v_\theta(x(t),t|\xr)
     \qquad \text{with} \qquad x(1) = r \sim \normal(0,1) \; .
\end{align}
We encode the velocity field using a transformer (TraCFM)~\cite{Butter:2023fov,Huetsch:2024quz}. The encoder processes the reco-level event and the decoder processes $x(t)$, in both cases with one token per feature. The input embeddings include a positional encoding of the features, and the decoder embedding additionally includes the time $t$. We use self-attention for the encoder and decoder separately to capture correlations between the respective features, and cross-attention to condition the decoder on the reco-level event. Each decoder output token then represents the velocity of its corresponding feature of $x(t)$.

\subsection{SM Higgs signal}
\label{sec:wbf_sig}

We first unfold the Standard Model signal. At parton level, the final state consists of the two tagging quarks, two leptons from the Higgs decay, and two neutrinos. The tagging-jet kinematics and the Higgs decay kinematics approximately factorize, because the scalar Higgs propagator carries no spin correlation. The only link between the two is the Higgs recoil,
\begin{alignat}{7}
 \vec{p}_{T,H} &=  - \vec{p}_{T,q_1} - \vec{p}_{T,q_2} \qqquad &&\text{(parton level)} \notag \\
 \vec{p}_{T,H} &\approx  - \vec{p}_{T,j_1} - \vec{p}_{T,j_2} \qqquad &&\text{(particle/reco level)}\; .
 \label{eq:higgs_propagator}
\end{alignat}
For the background suppression the reconstruction of the Higgs decay through the transverse mass is crucial. In contrast, for a CP analysis, the optimal observable~\cite{Brehmer:2017lrt,Butter:2021rvz} is the signed azimuthal angle between the two tagging jets~\cite{Plehn:2001nj,Hankele:2006ma},
\begin{align}
 \Delta\phi_{jj} = \phi_{j_1} - \phi_{j_2}
 \qquad \text{with} \qquad \eta_{j_1} > \eta_{j_2} \; .
\end{align}
Large jet transverse momenta and a wide rapidity gap reflect the characteristic signal's WBF topology. 

Whereas Eq.\eqref{eq:higgs_propagator} holds exactly at parton level, it is only approximate at reco level. Moreover, each quark has a fixed mass, whereas the jet masses vary. Both effects give the jets more degrees of freedom than the quarks.  We label the two quarks from the hard process by first ordering the tagging jets according to their transverse momenta and then choosing the labeling of the quarks that minimize 
\begin{align}
    \Delta R(q_{1,2},j_1) + \Delta R(q_{2,1},j_2)\; .
\end{align}
However, the generative unfolding pipeline does not depend on that choice as we unfold the full phase space rather than individual particles.

\subsubsection*{Naive phase space parameterization}

We first unfold events parameterized naively by the visible final state kinematics,
\begin{align}
    \Big\{ \; 
     (p_T, \phi, \eta)_{\ell_{1,2}},\,
     (p_T, \phi, \eta)_{q_{1,2}}
     \Big\} \; .
     \label{eq:naive_param}
\end{align}
The missing transverse momentum at parton level is obtained from the exact momentum balance between the Higgs system and the quarks
\begin{align}
    -\vec{p}_{T,\text{miss}} = \vec{p}_{T,\ell_1} +
    \vec{p}_{T,\ell_2} + \vec{p}_{T,q_1} + \vec{p}_{T,q_2}
     \label{eq:momentum_balance} \;.
\end{align}
Since the longitudinal momenta of the neutrinos are unknown, we cannot reconstruct the Higgs mass. Instead, we use the transverse mass
\begin{align}
     \label{eq:transverse_mass}
    m^2_{T} &= (E_{T,\text{miss}} + E_{T,\ell\ell})^2 - (\vec{p}_{T, \text{miss}} + \vec{p}_{T, \ell \ell})^2 \\
    &= m_{\ell\ell}^2 + m^2_{\text{miss}} + 2(E_{T, \ell\ell}\ E_{T, \text{miss}} - \vec{p}_{T, \ell\ell}\cdot \vec{p}_{T, \text{miss}})
    \qquad \text{with} \qquad E^2_{T,i} = p^2_{T,i} + m^2_{i} \quad (i = \ell\ell, \text{miss}) \; . \notag
\end{align}
Setting $m_{\text{miss}}=0$ guarantees that the transverse mass is bounded by the Higgs mass and improves the mass description~\cite{Barr:2009mx,Nath:2010zj}. An alternative choice is $m_{\text{miss}}=m_{\ell\ell}$, which is exact at the $WW$ threshold, but unbounded from above. We find no substantial difference between these choices in the unfolding and choose $m_{\text{miss}}=0$ throughout.

\begin{figure}[t]
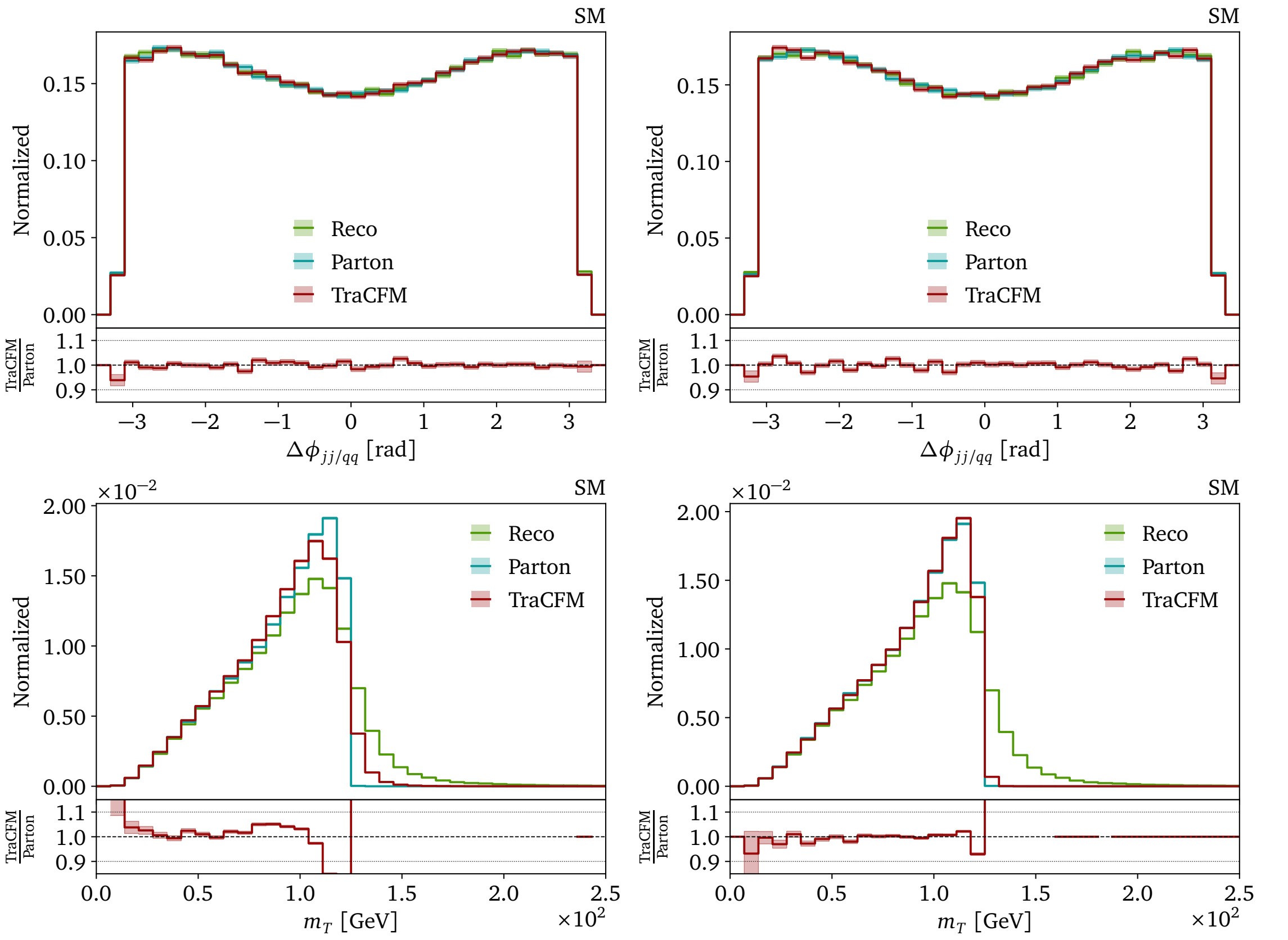

    \includegraphics[width=0.49\textwidth, page=21]{figs/signal-naive/parton-observables.pdf}
    \includegraphics[width=0.49\textwidth, page=22]{figs/signal-mt/parton-observables.pdf}\\
    \includegraphics[width=0.49\textwidth, page=24]{figs/signal-naive/parton-observables.pdf}
    \includegraphics[width=0.49\textwidth, page=29]{figs/signal-mt/parton-observables.pdf}
    \caption{Difference in signal unfolding performance between the naive parametrization (left) and the transverse mass parameterizations (right) for the TraCFM network.}
    \label{fig:signal-comparison}
\end{figure}

In Fig.~\ref{fig:signal-comparison} (left panels) we show the unfolded azimuthal angle between the tagging quarks $\Delta\phi_{qq}$ and the transverse mass of the Higgs decay products. While the unfolded $\Delta \phi_{qq}$ reproduces the SM truth reasonably well, the unfolded transverse mass largely misses the sharp edge.

\subsubsection*{Transverse mass parameterization}
\label{sec:mt_param_main}

The phase space parametrization defines which features the TraCFM network encodes directly versus through correlations. This allows us to improve the precision of challenging observables by including them explicitly~\cite{Ackerschott:2023nax,Huetsch:2024quz, Favaro:2025psi}. We define the transverse mass parameterization for parton-level events as the set of features
\begin{align}
    \Big\{ \; 
     (p_T, \phi, \eta)_{\ell_1},\;
     m_T,\;
     \Delta\eta_{\ell\ell},\;
     \Delta\phi_{\ell\ell},\;
     \Delta\phi_{\ell \ell,\text{miss}},\;
     r_{\ell \ell},\;
     \Delta\eta_{qq},\;
     \bar\eta_{q},\;
     H_{T,qq},\;
     \nu
     \Big\} 
\end{align}
with the simple definitions
\begin{alignat}{9}
     \Delta\eta_{\ell\ell} &= \eta_{\ell_1} - \eta_{\ell_2} 
     &\qqquad
     \Delta\phi_{\ell\ell} &= \phi_{\ell_1} - \phi_{\ell_2} 
     &\qqquad
     \Delta\phi_{\ell \ell,\text{miss}} &= \phi_{\ell\ell} - \phi_\text{miss} 
     &\qqquad
     r_{\ell\ell} &= \frac{m_{\ell\ell}}{m_T} < 1 \notag \\
     \Delta\eta_{qq} &= \eta_{q_1} - \eta_{q_2} 
     &
     \bar\eta_{q} &= \frac{\eta_{q_1} + \eta_{q_2}}{2} 
     &
     H_{T,qq} &= p_{T,q_1} + p_{T,q_2}\; ,
     \label{eq:transverse_mass_param}
\end{alignat}
and $\nu$ being the eccentric anomaly of an ellipse with major axis $H_{T,qq}$. We describe this basis in detail in App.~\ref{sec:appendix_mt_param}. The explicit transverse mass and the dilepton mass ratio $r_{\ell\ell}$ encode the dilepton mass as in turn
\begin{align}
    p_{T,\ell_2} = \frac{m^{2}_{\ell\ell}}{2p_{T,\ell_1}(\cosh{\Delta\eta_{\ell\ell}} - \cos{\Delta\phi_{\ell\ell}})}\; .
    \label{eq:ptl2_mt}
\end{align}
The transverse mass $m_T$ and $\Delta\phi_{\ell \ell,\text{miss}}$ determine the missing transverse momentum and, with it, the transverse Higgs recoil $\vec{p}_{T,H}$. Given $H_{T,qq}$, the transverse momentum of one of the quarks lies on an ellipse with foci at the endpoints of $-\vec{p}_{T,H}$, and major axis $H_{T,qq}$, and $\nu$ is its eccentric anomaly, see Fig.~\ref{fig:ellipse} in App.~\ref{sec:appendix_mt_param}. This parameterization applies only to parton-level events. The reco-level input is
\begin{align}
    \Big\{\;
    (p_T,\phi,\eta)_{\ell_1},\;
    p_{T,\ell\ell},\;
    m_T,\;
    \Delta\eta_{\ell\ell},\;
    \Delta\phi_{\ell\ell},\;
    \Delta\phi_{\ell\ell,\text{miss}},\;
    (p_T,\phi,\eta,m)_{j_{1,2}}\;
    \Big \}\; .
\label{eq:input_para}
\end{align}
It has four extra degrees of freedom compared to parton level, coming from the jet mass and Eq.\eqref{eq:higgs_propagator}. 

\begin{figure}[b!]
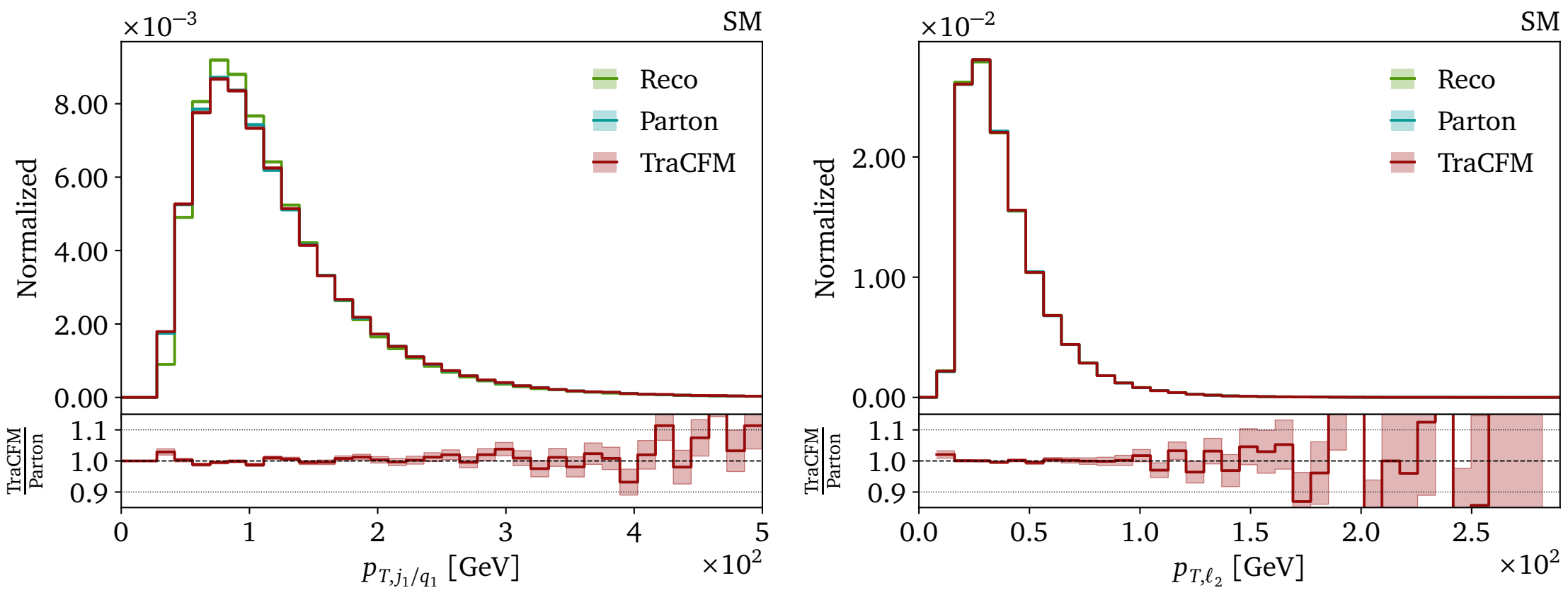

    \includegraphics[width=0.49\textwidth, page=7]{figs/signal-mt/parton-observables.pdf}
    \includegraphics[width=0.49\textwidth, page=4]{figs/signal-mt/parton-observables.pdf} 
    \caption{Unfolded signal distributions of the quark (left) and lepton (right) transverse momentum at parton level, using the transverse mass parameterization.}
    \label{fig:signal-mt}
\end{figure}

In App.~\ref{sec:appendix_mt_param}, we show that every unfolded event satisfies transverse momentum conservation and passes the quark selection in Tab.~\ref{tab:selection} by construction. In our forward simulation, each parton-level lepton has a one-to-one and precise correspondence with a reco-level lepton. We therefore do not unfold the lepton kinematics $x_{\text{part},\ell}$ directly, but their shift with respect to the reco-level values,
\begin{align}
    \label{eq:delta_lepton}
    \delta_\ell \equiv x_{\text{part},\ell} - x_{\text{reco},\ell} 
     \qquad \Leftrightarrow \qquad 
      x_{\text{part},\ell} = \delta_\ell  + x_{\text{reco},\ell} \;.
\end{align}
Whereas the narrow distribution $p(x_{\text{part},\ell}|\xr)$ moves with the reco-level event and consequently covers the whole phase space, $p(\delta_{\ell}|\xr)$ stays centered close to zero. The width, set by the detector resolution, changes only slowly with $\xr$. In practice, the reco-level and parton-level features are all preprocessed by invertible functions and we compute the shift in Eq.\eqref{eq:delta_lepton} between preprocessed features. For quarks, we lack such a one-to-one correspondence with the jets, and jets have intrinsically more degrees of freedom than quarks. We therefore unfold the quark features directly.

We show the unfolded distributions using this alternative phase space parametrization in the right panels of Fig.~\ref{fig:signal-comparison}. The unfolded transverse mass improves noticeably over the whole distribution and in particular, at the sharp kinematic edge. The other observables, such as $\Delta\phi_{jj}$, retain their precision. In Fig.~\ref{fig:signal-mt} we illustrate the unfolding performance for the transverse momenta of the leading quark and the subleading lepton, both observables no longer unfolded directly, but derived from unfolded features.

\subsection{SMEFT Higgs signal}
\label{sec:wbf_eft}

\begin{figure}[t!]
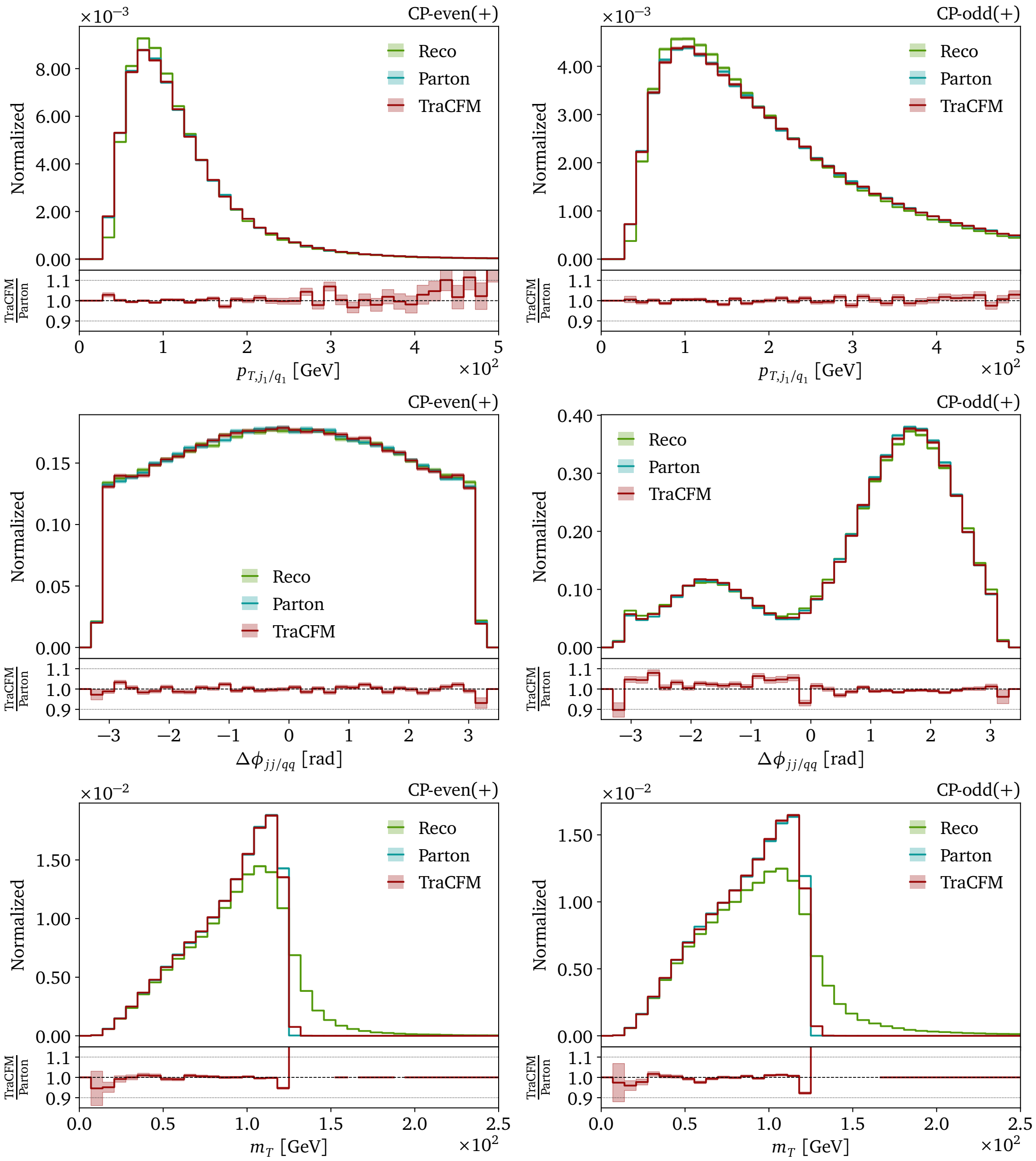

    \includegraphics[width=0.48\textwidth, page=7]{figs/even-mt/parton-observables.pdf}
    \includegraphics[width=0.48\textwidth, page=7]{figs/odd-mt/parton-observables.pdf} \\
    \includegraphics[width=0.48\textwidth, page=23]{figs/even-mt/parton-observables.pdf}
    \includegraphics[width=0.48\textwidth, page=23]{figs/odd-mt/parton-observables.pdf} \\
    \includegraphics[width=0.48\textwidth, page=29]{figs/even-mt/parton-observables.pdf}
    \includegraphics[width=0.48\textwidth, page=29]{figs/odd-mt/parton-observables.pdf}
    \caption{Unfolded CP-even (left) and CP-odd (right) distributions at parton level using the transverse mass parameterization, with the TraCFM network trained consistently on the respective SMEFT datasets.}
    \label{fig:even-odd-mt}
\end{figure}

Now that we know that TraCFM can unfold the WBF phase space and, in particular, the CP observable $\Delta \phi_{jj}$, we study a potential bias from the training data. We extend the Lagrangian to include the CP-even operator $\ope_{HW}$ and the CP-odd operator $\ope_{\widetilde{HW}}$. Both operators modify the Lorentz structure of the $HWW$ vertex and affect the kinematics of the tagging jets and of the Higgs decay products. First, we evaluate the unfolding performance after training on simulated SMEFT event pairs including 
\begin{align}
 \frac{c_{HW}}{\Lambda^2} &= \frac{0.5}{\tev^2} \qqqquad \text{CP-even($+$)} \notag \\
 \frac{c_{\widetilde{HW}}}{\Lambda^2} &= \frac{5.0}{\tev^2} \qqqquad \text{CP-odd($+$)} \; .
\label{eq:def_smeft}
\end{align}
We then unfold a test sample with the same SMEFT contributions. In Fig.~\ref{fig:even-odd-mt} we find percent-level precision across phase space, in particular in the jet observables and the transverse mass. 

We emphasize the impact of the CP-even and CP-odd operators on the azimuthal angle difference between the tagging jets, as well as the considerable broadening of the transverse momenta of the tagging jets for the CP-odd operator. The effect on the kinematics of the Higgs decay products is relatively mild and restricted to a broadening of the transverse mass. Independent of the SMEFT operator, the unfolding yields reliable results with high precision when the training data and the observed data are consistent.

\subsubsection*{Unfolding SMEFT under SM assumption}

\begin{figure}[t!]
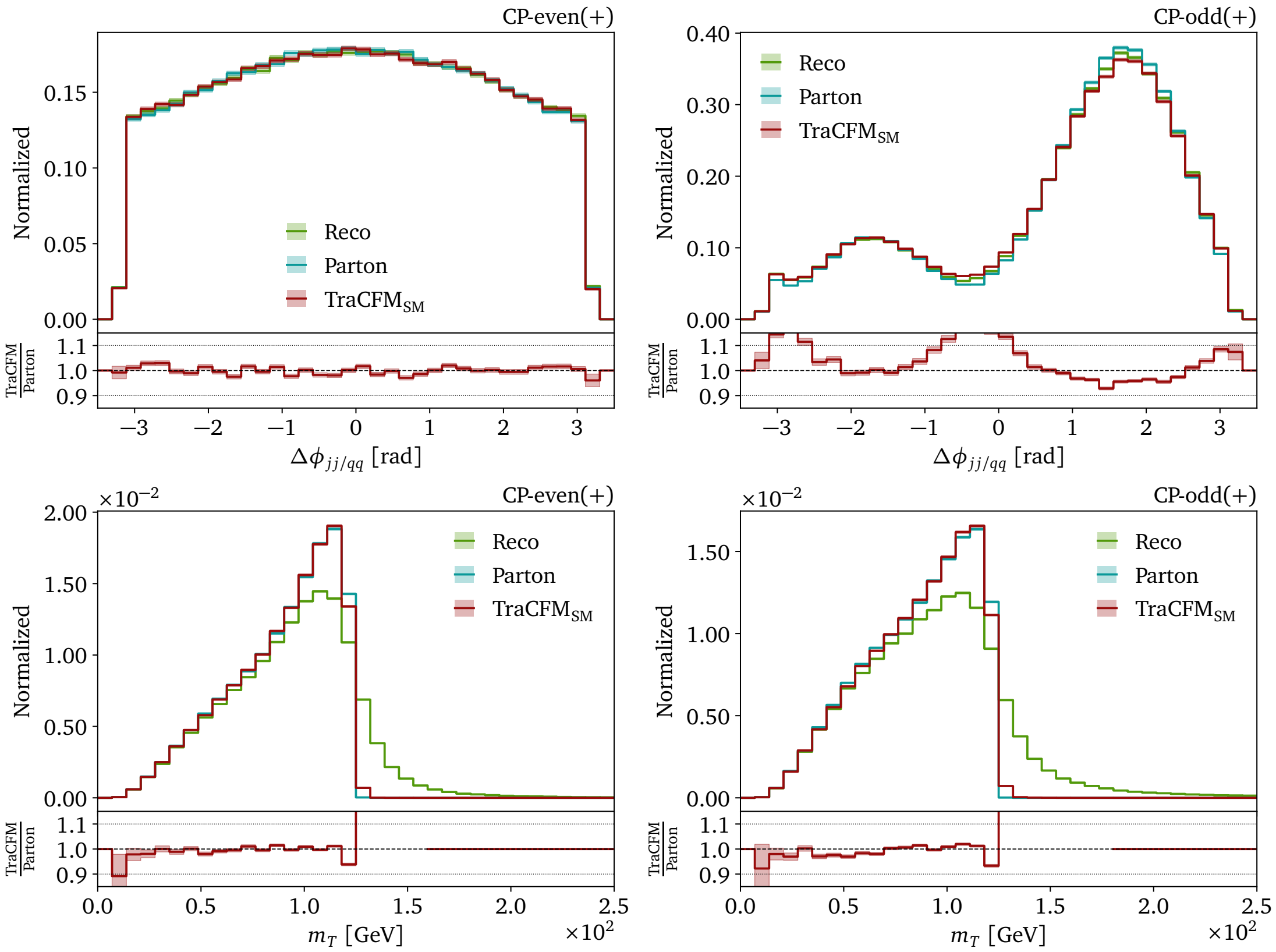

    \includegraphics[width=0.48\textwidth, page=23]{figs/signal-mt/eval_even-parton-observables.pdf}
    \includegraphics[width=0.48\textwidth, page=23]{figs/signal-mt/eval_odd-parton-observables.pdf} \\
    \includegraphics[width=0.48\textwidth, page=29]{figs/signal-mt/eval_even-parton-observables.pdf}
    \includegraphics[width=0.48\textwidth, page=29]{figs/signal-mt/eval_odd-parton-observables.pdf}
    \caption{Unfolded CP-even (left) and CP-odd (right) distributions at parton level, with the TraCFM network trained on SM events.}
    \label{fig:train_sm-eval_cpeven-odd}
\end{figure}

To identify SMEFT contributions in the unfolded data, the prior dependence of Eq.\eqref{eq:prior_dependence} has to be smaller than the effects of dimension-6 operators. We test this by training the network on the SM sample and applying it to the CP-even and CP-odd samples, shown in Fig.~\ref{fig:train_sm-eval_cpeven-odd}. For the CP-even sample, there is no visible deviation between the unfolded distributions and the parton-level truth, as the smaller kinematic SMEFT effects in combination with the good lepton and jet resolution suppress the bias. In particular, the reco- and parton-level distributions of $\Delta \phi_{jj/qq}$ are almost identical, so that the unfolding hardly has to correct this observable. 

For the CP-odd sample, we observe a bias of up to 15\% in the azimuthal angle difference at the minima around $\pm 3$ and $-0.5$~rad. In the remaining observables we only find a very mild bias towards smaller transverse momentum of the leading jet. Compared to the effect of the CP-odd operator itself, the unfolding bias is small.

\subsubsection*{Unfolding SMEFT under conditional assumptions}

\begin{figure}[t!]
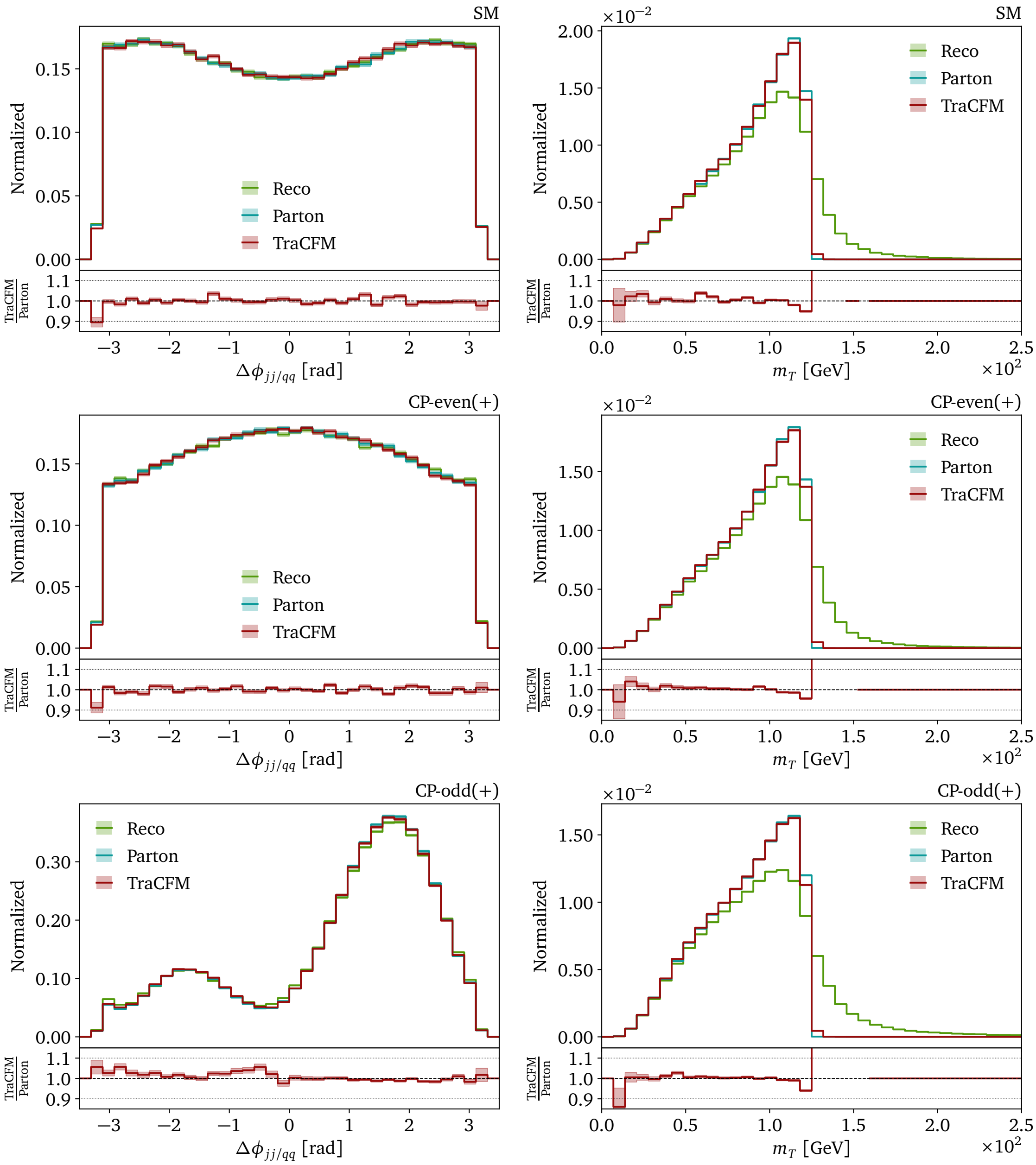

    \includegraphics[width=0.49\textwidth, page=23]{figs/allsamples/eval_sm-observables.pdf}
    \includegraphics[width=0.49\textwidth, page=29]{figs/allsamples/eval_sm-observables.pdf} \\
    \includegraphics[width=0.49\textwidth, page=23]{figs/allsamples/eval_cpeven_pos-observables.pdf}
    \includegraphics[width=0.49\textwidth, page=29]{figs/allsamples/eval_cpeven_pos-observables.pdf}\\
    \includegraphics[width=0.49\textwidth, page=23]{figs/allsamples/eval_cpodd_pos-observables.pdf}
    \includegraphics[width=0.49\textwidth, page=29]{figs/allsamples/eval_cpodd_pos-observables.pdf}
    \caption{Unfolded SM (top), CP-even (center), and CP-odd (bottom) distributions at parton level, with the TraCFM network trained on SM and SMEFT samples combined.}
    \label{fig:train_all-eval_separate}
\end{figure}

The standard approach to reducing the prior dependence is to include alternative hypotheses in the training data, such that it covers all potential unfolding outcomes and consequentially broadens the prior distribution $p_\text{sim}(\xp)$ in Eq.\eqref{eq:prior_dependence}. We train a single TraCFM network on the SM sample and the four SMEFT samples defined in Eq.\eqref{eq:def_smeft} with the label
\begin{align}
\lambda \in 
\Big\{ \; 
\text{SM}, 
\text{CP-even($\pm$)},
\text{CP-odd($\pm$)} \; 
\Big\} \; ,
\end{align}
using equal numbers of events per hypothesis. Each training event carries its label, which enters the network through a linear embedding layer, see Tab.~\ref{tab:hyper_unf} in App.~\ref{sec:hyperparams}. All other network parameters are shared between the hypotheses, so the network learns 
\begin{align}
    \pmd(\xp|\xr,\lambda)\approx p_{\text{sim}, \lambda}(\xp|\xr)\; .
\end{align}
The common mappings between jets, leptons, and missing transverse momentum at reco level and at parton level are learned from all events, while the label generates  changes in the parton-level distributions, for instance, different correlations between the tagging quarks.

In Fig.~\ref{fig:train_all-eval_separate} we show that the unfolded SM, CP-even, and CP-odd data agrees with the assumed  parton-level truth at the percent level. Unfolding each sample with its corresponding label is just a closure test, to confirm that the network learns all posteriors. For the azimuthal angle difference $\Delta\phi_{jj}$, computed from the unfolded quark momenta, we find not only excellent agreement for all underlying models, but an improvement over single-sample training. The transverse mass of the Higgs system, unfolded directly, also reaches at least the same accuracy as training per hypothesis. This shows that the increased number of combined training events improves the general accuracy.

\subsection{Conditional unfolding envelope}
\label{sec:wbf_env}

\begin{figure}[t!]
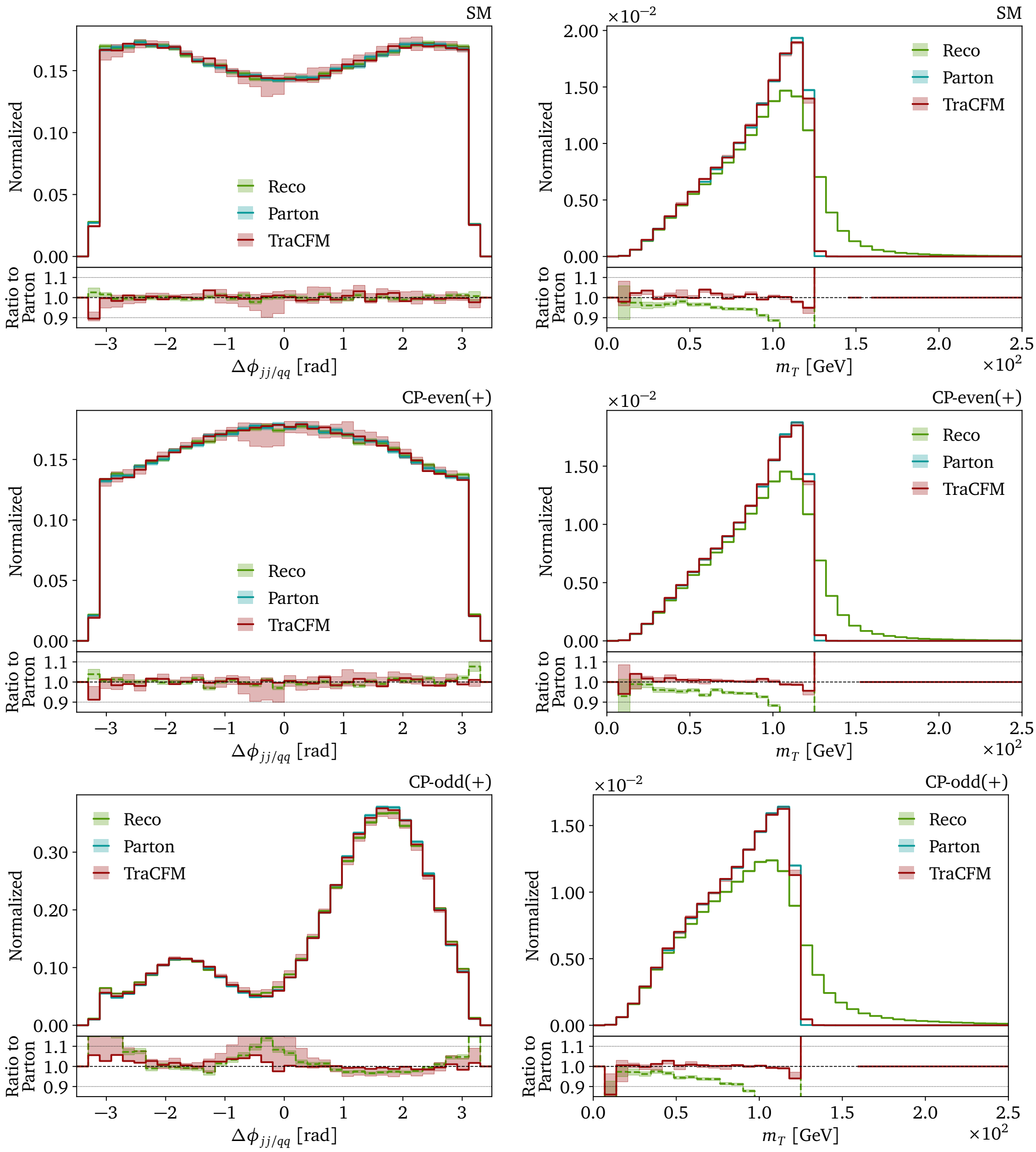

    \includegraphics[width=0.49\textwidth, page=23]{figs/allsamples/eval_sm_wronglabels_spread_envelope.pdf}
    \includegraphics[width=0.49\textwidth, page=29]{figs/allsamples/eval_sm_wronglabels_spread_envelope.pdf}
    \includegraphics[width=0.49\textwidth, page=23]{figs/allsamples/eval_cpeven_pos_wronglabels_spread_envelope.pdf}
    \includegraphics[width=0.49\textwidth, page=29]{figs/allsamples/eval_cpeven_pos_wronglabels_spread_envelope.pdf}
    \includegraphics[width=0.49\textwidth, page=23]{figs/allsamples/eval_cpodd_pos_wronglabels_spread_envelope.pdf}
    \includegraphics[width=0.49\textwidth, page=29]{figs/allsamples/eval_cpodd_pos_wronglabels_spread_envelope.pdf}
    \caption{Unfolded SM, CP-even, and  CP-odd distributions at parton level, with the TraCFM network trained conditionally. We indicate the bin-wise range of the five labels in red.}
    \label{fig:trainSMEvenOdd-unfoldSMEvenOdd_spread}
\end{figure}

When unfolding measured data, the label $\lambda$ is a discrete nuisance parameter, turning Eq.\eqref{eq:def_unfold} into
\begin{align}
    p_\text{unfold}(\xp|\lambda) = \int\d \xr\ \pmd(\xp|\xr,\lambda) \ \pd(\xr)\; .
\end{align}
To quantify the systematic uncertainty we unfold each set of reco-level events with all five labels. For each bin $i$ of an observable, this gives the normalized bin contents $n_i(\lambda)$. As for a nuisance parameter with a uniform prior, the uncertainty band is the per-bin envelope
\begin{align}
    \Big[ \min_\lambda n_i(\lambda),\ \max_\lambda n_i(\lambda) \Big] \; .
\end{align}
This choice is naturally conservative, as the end points per bin can come from all labels. Since all labels unfold the same reco-level events, the statistical fluctuations of the different $n_i(\lambda)$ are fully correlated, so the range only reflects a potential training instability and the training data bias. A narrow envelope implies that the unfolded distribution is determined well by the reco-level event, and we expect the statistical uncertainty of the events to become the lower limit on the systematics. 

In Fig.~\ref{fig:trainSMEvenOdd-unfoldSMEvenOdd_spread} we show the envelopes for the SM, CP-even, and CP-odd samples at reco level, unfolded with the network trained conditionally on all five hypotheses. For the Higgs decay observables, including the transverse mass, the envelope is comparable to the statistical uncertainty, so even the worst-label prediction is compatible with the truth.

For the jet distributions, especially $\Delta\phi_{jj}$, the envelope covers around three times the statistical uncertainty. The reason is that the tagging jet kinematics are most affected by the dimension-6 operators. One effect of the envelope construction is that the low training statistics for the CP-odd case around $\Delta\phi_{jj} = 0$ translates into a large systematic when we unfold SM and CP-even data.
This is confirmed by Fig.~\ref{fig:label_contributions_sm_cpevenpos_cpoddpos}, in App.~\ref{sec:wbf_env_contributions}, where most of the width of the envelope is driven by the CP-odd($+$) deviation with respect to the other four hypotheses.

This means the label dependence is not uniform across observables, it is large only where the reco-level event does not fully constrain the parton-level kinematics, as for the tagging jets, and it stays at the level of the statistical uncertainty elsewhere. This suggests a strategy for an actual measurement, performing an inclusive training on a set of physics hypotheses, unfold the measured data with each label, and report the envelope as the remaining simulation dependence. This does not require knowing which hypothesis is realized in the data, provided it lies within the span of the training hypotheses.

\section{Background removal}
\label{sec:back}
The $t\bar{t}$ sample is generated without the $b$-veto or the central jet veto, to provide sufficient training statistics. We instead account for the vetoes through a global rescaling of the $t\bar{t}$ rate by the combined veto efficiency $\epsilon_\text{veto} = 0.09$, extracted from the cut flow in Tab.~\ref{tab:cutflow_parton}, so that $\alpha_{t\bar{t}}$ corresponds to the post-veto rate. This assumes that the vetoes do not significantly change the kinematic distributions seen by the classifier. This is the case by construction, since the classifier only sees the two tagging jets and no $b$-tagging information. The outside lepton veto changes the $t\bar{t}$ kinematics, so we apply it explicitly at reco and truth levels. Even after a tight selection, the top background rate remains much larger than the WBF Higgs signal rate, and even tighter cuts are clearly undesirable for unfolding. We therefore remove backgrounds statistically through classifier reweighting~\cite{Andreassen:2021zzk,Butter:2025mek}. Using the relative signal and background rates~\cite{ATLAS:2023pwa}
\begin{align}
\alpha_{Hjj} &= 0.144 \notag \\
\alpha_{WWjj} &= 0.320 \notag \\
\alpha_{t\bar{t}} &= 0.536  \qquad \mwith \qquad \sum_i \ai = 1\; ,
\end{align}
the normalized signal, background, and combined data densities are given by 
%
\begin{align}
  \pd(x) &= \alpha_{Hjj}\,\ps(x) + \alpha_\text{bkg}\,\pbg(x) \notag \\
  \pbg(x) &= \frac{\alpha_{WWjj}\, p_{WWjj}(x) + \alpha_{t\bar{t}}\, p_{t\bar{t}}(x)}{\alpha_\text{bkg}} \qquad
  \mwith \qquad \alpha_\text{bkg} = \alpha_{WWjj} + \alpha_{t\bar{t}}\; ,
\end{align}
in terms of the normalized distributions $p_{WWjj}$ and $p_{t\bar{t}}$. 

A Monte-Carlo background estimate $\pbg^{\text{MC}}(x)$ approximates $\pbg(x)$ up to statistical fluctuations. We split our simulated background sample into halves, one included in $\pd(x)$ to mimic the true background and the other representing $\pbg^{\text{MC}}(x)$. Our goal is to learn event weights $w(x)$ such that
\begin{align}
        w(x) 
        &= \frac{\alpha_\text{bkg}\,\pbg^{\text{MC}}(x)}{\pd(x)} \notag \\
    \Leftrightarrow \qquad 
    [1-w(x)]\,\pd(x) 
    &=
    \pd(x)-\alpha_\text{bkg}\,\pbg^{\text{MC}}(x) 
    \;\overset{\text{MC bkg}}{=}\; 
    \alpha_{Hjj}\ps(x) \; .
    \label{eq:goal}
\end{align}
To achieve this, one could train a binary classifier, where the training events carry a label $y$ and a weight $\omega$
\begin{itemize}
  \item $y=0$: events $x\sim\pd(x)$, with positive weights, normalized to unity;
  \item $y=1$: events $x\sim\pd(x)$, with positive weights normalized to unity, together with events $x\sim\pbg^{\text{MC}}(x)$, with negative weights summing to $-\alpha_\text{bkg}$ \;.
\end{itemize}
The output of this classifier indeed approximates the target likelihood ratio from Eq.\eqref{eq:goal},
\begin{align}
    \frac{C(x)}{1-C(x)} \simeq \frac{\pd(x) - \alpha_\text{bkg}\,\pbg^{\text{MC}}(x)}{\pd(x)} \; .
    \label{eq:reweight1}
\end{align}
A weighted cross-entropy loss extracting $\pd(x) - \alpha_\text{bkg}\,\pbg^{\text{MC}}(x)$ from mixed-sign events implicitly requires $\alpha_\text{bkg}\,\pbg^{\text{MC}}(x)\leq\pd(x)$. Statistical fluctuations violate this condition locally, so if the network manages to resolve these fluctuations, the training is ill-posed, the output will be driven to zero, and the loss will become unbounded from below. 

Instead, we turn to an unweighted classifier learning the ratio of the two normalized densities
\begin{align}
   \frac{C(x)}{1-C(x)} \simeq \frac{\pbg^{\text{MC}}(x)}{\pd(x)} \; .
\end{align}
We re-scale it to obtain the background weight defined in Eq.\eqref{eq:goal}
\begin{align}
        w(x) 
        = \alpha_\text{bkg}\frac{C(x)}{1-C(x)}
        \simeq \frac{\alpha_\text{bkg}\,\pbg^{\text{MC}}(x)}{\pd(x)} \; ,
        \label{eq:reweight2}
\end{align}
and then use it to reweight via
\begin{align}
        1-w(x) 
        \simeq \frac{\pd(x) - \alpha_\text{bkg}\,\pbg^{\text{MC}}(x)}{\pd(x)}
        \simeq \frac{\alpha_{Hjj}\,\ps(x)}{\pd(x)}\; ,
        \label{eq:reweight2.1}
\end{align}
just as in Eq.\eqref{eq:goal}.

Comparing Eqs.\eqref{eq:reweight1} and~\eqref{eq:reweight2.1}, the training strategies are different. In the second strategy, the class densities are non-negative by construction, and the optimization is well-posed. The negative weights have been moved from the training to the reweighting. The extracted weight $w(x)$ is no longer bounded by one, in background-dominated regions it fluctuates around it. After training, around 7\% of events have $w(x)>1$. We keep them, as removing them or clipping their weights would bias the subtracted background.

\subsubsection*{Classifier training}

\begin{figure}[b!]
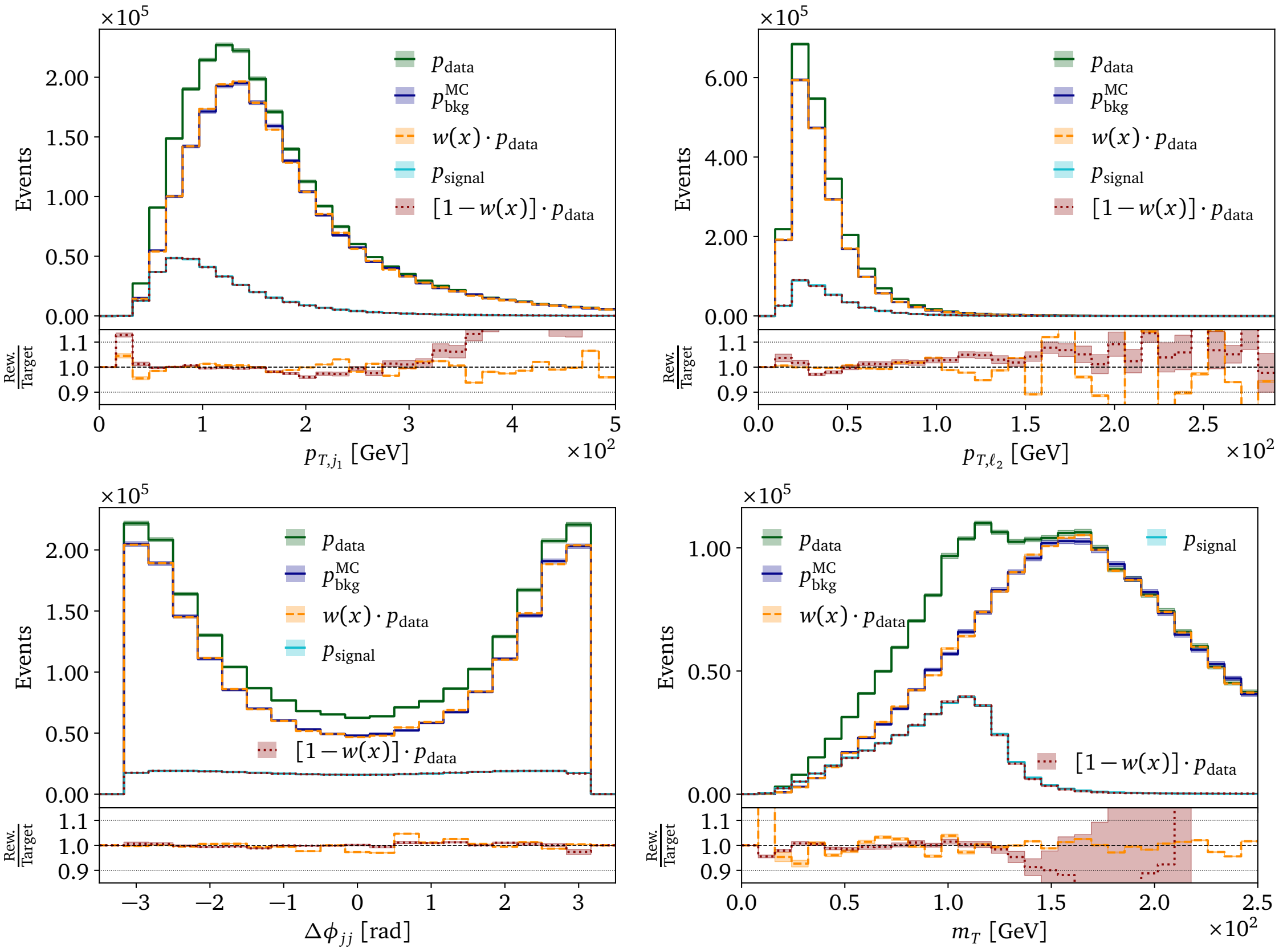

    \includegraphics[width=0.49\textwidth, page=7]{figs/bkg-subtraction/member-normalized/rew-reco-observables.pdf}
    \includegraphics[width=0.49\textwidth, page=4]{figs/bkg-subtraction/member-normalized/rew-reco-observables.pdf}
    \includegraphics[width=0.49\textwidth, page=24]{figs/bkg-subtraction/member-normalized/rew-reco-observables.pdf}
    \includegraphics[width=0.49\textwidth, page=27]{figs/bkg-subtraction/member-normalized/rew-reco-observables.pdf}
    \caption{Background removal at reco level.
    We show the data, the MC background estimate, and the true signal, together with the data reweighted with $w(x)$ and $1-w(x)$, Eqs.\eqref{eq:reweight2} and~\eqref{eq:reweight2.1}. The reweighted lines and bands are the mean and standard deviation over the five ensemble members. The lower panels show the ratio to the respective target.}
    \label{fig:class}
\end{figure}

We use the classifier inputs
\begin{align}
      \Big\{\;
      (p_T,\phi,\eta)_{\ell_{1,2}},\;
      (p_T,\phi,\eta,m)_{j_{1,2}},\;
      (p_T,\phi)_\text{miss},\;
      \Delta\phi_{\ell\ell},\;
      \Delta\eta_{\ell\ell},\;
      E_{T,\ell\ell},\;
      m_T,\;
      \Delta\eta_{jj},\;
      m_{jj},\;
      \Delta\phi_{jj}
      \;\Big\}\; ,
\end{align}
each mapped to an approximate Gaussian. The hyperparameters are listed in Tab.~\ref{tab:hyper_clf} in App.~\ref{sec:hyperparams}. Each background event is weighted with the rate $\alpha_i$ of its process, and the $t\bar{t}$ events additionally carry the weight from the reweighting to parton-level cuts described in Sec.~\ref{sec:sim}.

We train with $k$-folding, randomly splitting the dataset into three folds. Each event obtains its weight from classifiers trained only on the other two folds. The signal estimate is noisy, because it represents only about 14\% of the data and small deviations in the subtracted background translate into large relative errors on the signal. We train an ensemble of five classifiers with different random initializations and average the reweighted histograms over the ensemble. Each member $k$ subtracts a total background
\begin{align}
    B_k = \int \d x\, w_k(x)\, \pd(x)\; ,
\end{align}
which should be equal to $\alpha_\text{bkg}$ but, in practice, deviates from it by around $1\%$. This small deviation causes large biases in the resulting reweighted signal distribution. For our proof of principle, we therefore assume the total background rate as known and fix the normalization of each member via 
\begin{align}
    w_k(x)\rightarrow \frac{\alpha_\text{bkg}}{B_k}\, w_k(x)\; .
\end{align}
Now, each member subtracts exactly the background yield, and the spread over the members measures the remaining uncertainty in the shape of the subtracted background.

\subsubsection*{Unfolding while removing backgrounds}

\begin{figure}[t]
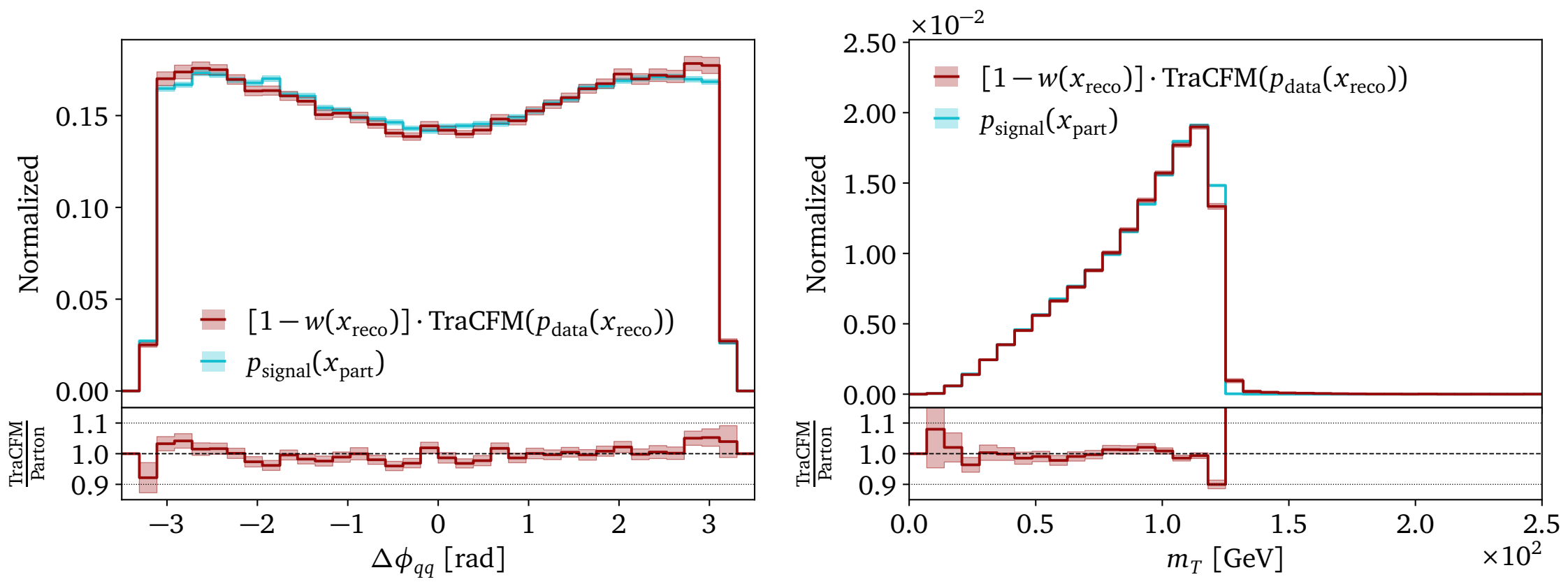

    \includegraphics[width=0.49\textwidth, page=23]{figs/bkg-subtraction/member-normalized/unfolded_data-parton-observables}
    \includegraphics[width=0.49\textwidth, page=29]{figs/bkg-subtraction/member-normalized/unfolded_data-parton-observables}
    \caption{Unfolded background-removed data (red) and parton-level signal truth (blue), for $\Delta\phi_{qq}$ (left) and $m_T$ (right). The band combines the ensemble spread with the bootstrap uncertainty.}
    \label{fig:unfolded_class}
\end{figure}

We show the results for the reweighted distributions at reco level in Fig.~\ref{fig:class}. Weighting data with $w(x)$  reproduces the MC background within 3\% in the bulk of the distributions; weighting it with $1-w(x)$ reproduces the true signal within 3-4\%. We obtain the signal estimate as the small difference between data and background, so large fluctuations appear in the background-dominated tails.

We then unfold the reweighted data with a TraCFM network trained on the SM signal. Every reco-level data event carries the weight $1-w(\xr)$. In Fig.~\ref{fig:unfolded_class} we compare the unfolding outcome with the parton-level truth of the signal events. The two agree at the percent level in the bulks of the distributions, with some deviations in the tails and towards the edges of the phase space. The uncertainty band combines the ensemble spread of the classifiers with the statistical uncertainty from 200 Poisson bootstrap replicas of the data. A signal event enters both the unfolded distribution and the truth with the same multiplicity, so its uncertainty cancels in the ratio, while the background events only enter the unfolded distribution.

\section{Parameter estimation of unfolded data}
\label{sec:para}

\begin{table}[b!]
    \centering
    \resizebox{\textwidth}{!}{
    \begin{tabular}{l cc cc cc}
    \toprule
     & \multicolumn{2}{c}{SM, $c_{\widetilde{HW}} = 0$} 
     & \multicolumn{2}{c}{CP-odd($+$), $c_{\widetilde{HW}} = 5$} 
     & \multicolumn{2}{c}{CP-odd($-$), $c_{\widetilde{HW}} = -2$} \\
    \cmidrule(lr){2-3} \cmidrule(lr){4-5} \cmidrule(lr){6-7}
     & $\hat{c}_{\widetilde{HW}}$ & 68\% CL 
     & $\hat{c}_{\widetilde{HW}}$ & 68\% CL 
     & $\hat{c}_{\widetilde{HW}}$ & 68\% CL \\
    \midrule
    SM-only TraCFM          &  $0.006$ & $[-1.20, 1.20]$ & $5.020$ & $[4.24, 5.78]$ & $-1.939$ & $[-3.06, -0.76]$ \\
    SM$+$EFT TraCFM (SM condition) &  $0.002$ & $[-1.20, 1.20]$ & $5.017$ & $[4.24, 5.78]$ & $-1.951$ & $[-3.06, -0.78]$ \\
    SM-only TraCFM, bkg-subtracted & $0.025$ & $[-1.18, 1.22]$ & \multicolumn{2}{c}{---} & \multicolumn{2}{c}{---} \\
    Reco level                   & $-0.006$ & $[-1.25, 1.23]$ & $4.998$ & $[4.22, 5.76]$ & $-2.001$ & $[-3.09, -0.82]$ \\
    \bottomrule
    \end{tabular}
    }
    \caption{Best-fit values and 68\% CL intervals of $c_{\widetilde{HW}}$, in units of $\text{TeV}^{-2}$, from the unfolded $\Delta\phi_{qq}$ and reco-level $\Delta\phi_{jj}$ distributions for $300~\text{fb}^{-1}$. The column headers give the pseudo-data and the true value of $c_{\widetilde{HW}}$. The background-subtracted fit uses SM signal with the $t\bar{t}$ and $WWjj$ backgrounds as pseudo-data, and is therefore only available for the SM hypothesis.}
    \label{tab:MLEvalues}
\end{table}

Finally, we test whether a simple inference or measurement on unfolded data recovers the correct physics parameters. We extract the CP-odd Wilson coefficient $c_{\widetilde{HW}}$ by fitting the unfolded signed $\Delta\phi_{qq}$ distribution as its optimal observable, and compare the result with a fit to corresponding reco-level observable. As the assumed truth value we use the signal-only SM and CP-odd($\pm$) event samples at reco level as pseudo-data, \ie 
\begin{align}
  \left( c_{HW} = 0, c_{\widetilde{HW}} = 0 \right) 
  \qqqquad
  \left( c_{HW} = 0, c_{\widetilde{HW}} = 5 \right) 
  \qqqquad 
  \left( c_{HW} = 0, c_{\widetilde{HW}} = -2 \right) \; .
\end{align}
For fits to the unfolded parton-level and reco-level data, we analyze bin-wise predictions $n_i$ as a function of $c_{\widetilde{HW}}$. All histograms are normalized to their respective generator-level cross sections and scaled to an integrated luminosity of $300\ \text{fb}^{-1}$. Since we generate the Higgs decay with both $W$ bosons close to their mass shell, our signal rate is smaller than in the ATLAS analysis~\cite{ATLAS:2025hki}, and the widths of our 68\% CL intervals are not directly comparable to the experimental results. For our truncated dimension-6 SMEFT modification, we know that the amplitude at each phase space point has the functional form~\cite{Butter:2016cvz}
\begin{align}
    n_i(c_{\widetilde{HW}}) = n_i^\text{SM}\Big (1 + A_i\ c_{\widetilde{HW}} + B_i\ c_{\widetilde{HW}}^2\Big )\; .
    \label{eq:bin_rates}
\end{align}
We derive this closed form in terms of $n_i^\text{SM}$, $A_i$, and $B_i$ by solving this equation for the binned SM and CP-odd($\pm$) distributions using Effective Lagrangian Morphing~\cite{Balasubramanian:2022vzv}. We then evaluate this closed form as a function of $c_{\widetilde{HW}}$ and compare each prediction to the pseudo-data through the binned Poisson likelihood
\begin{align}
    -\ln L(c_{\widetilde{HW}}) = \sum_{i=1}^{4} \Big[ n_i(c_{\widetilde{HW}}) - m_i \ln n_i(c_{\widetilde{HW}}) \Big] + \text{const} \; ,
\end{align}
where $m_i$ is the number of pseudo-data events in bin $i$ and $n_i$ is the corresponding prediction. Since every input event is unfolded with a single sampling, the bins fluctuate as independent Poisson variables. We extract the best-fit $\hat{c}_{\widetilde{HW}}$ by minimizing $- \ln L$ and the statistical uncertainties through a $68\%$ CL interval defined by $-\ln L < 1/2$.

\begin{figure}[t]
    \includegraphics[width=0.49\textwidth]{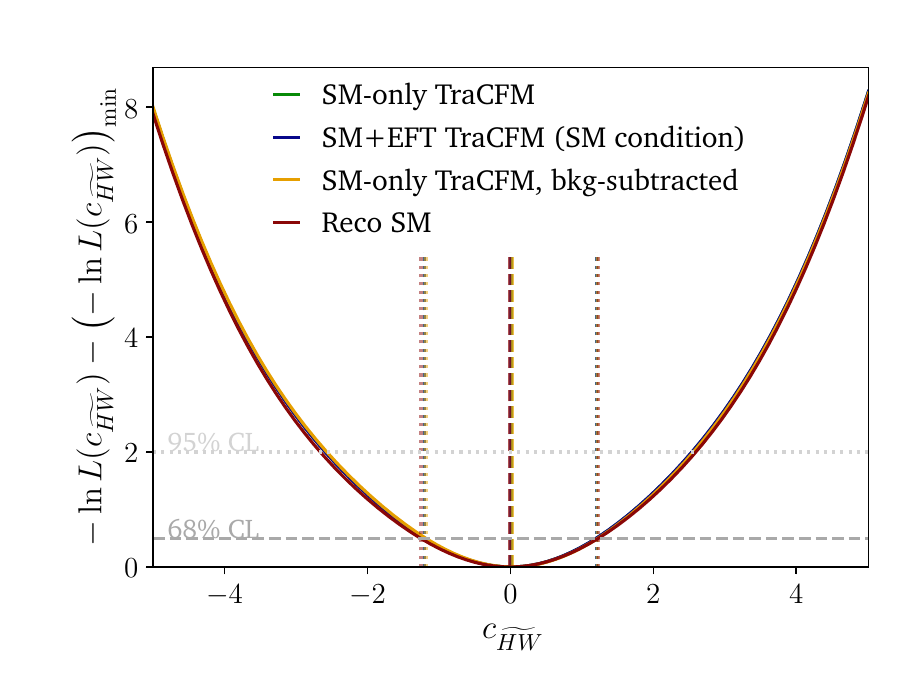}
    \includegraphics[width=0.49\textwidth]{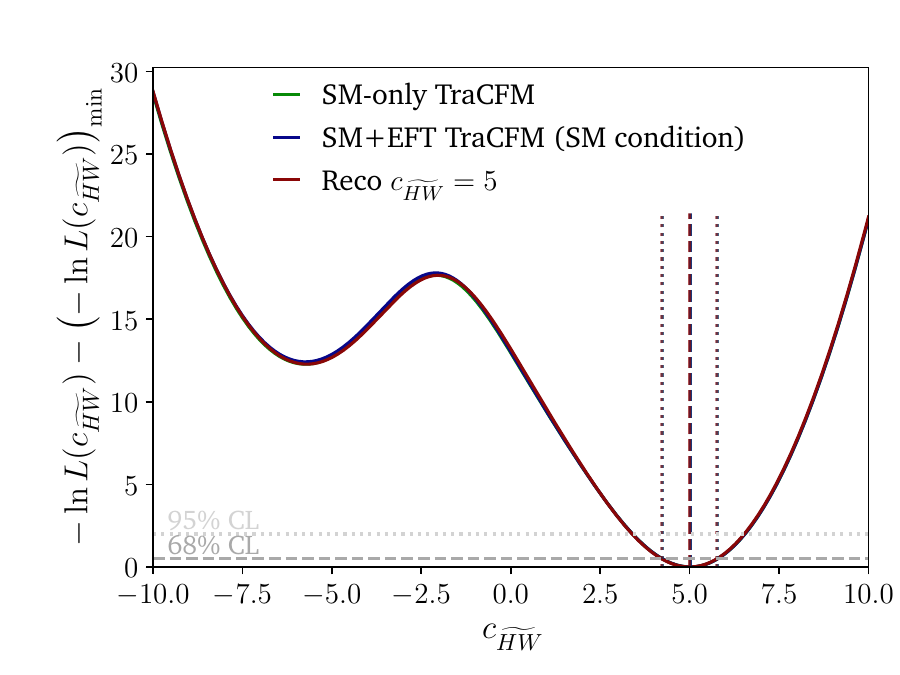} \\
    \includegraphics[width=0.49\textwidth]{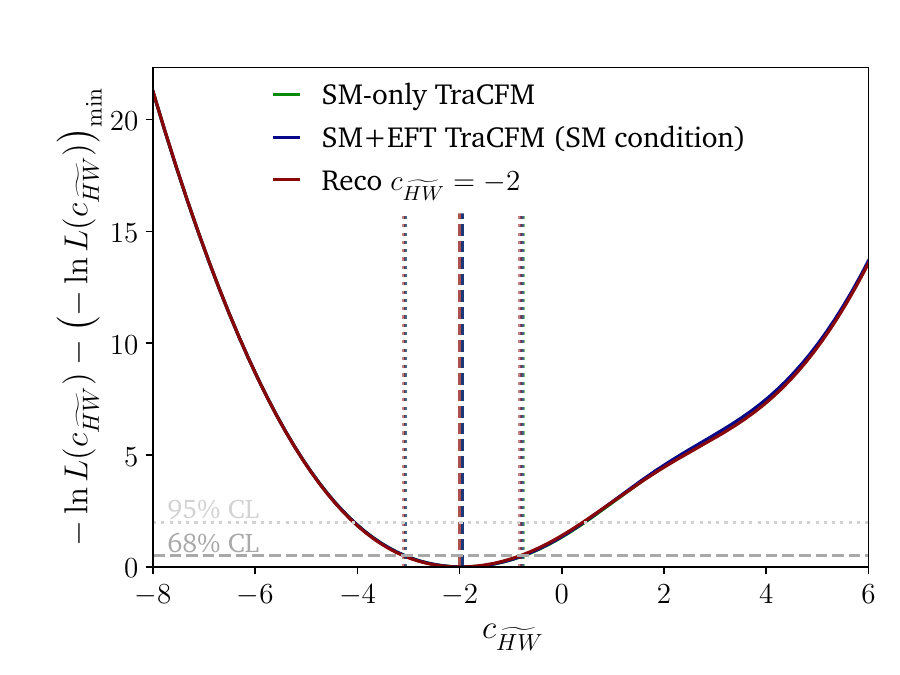}
    \includegraphics[width=0.49\textwidth]{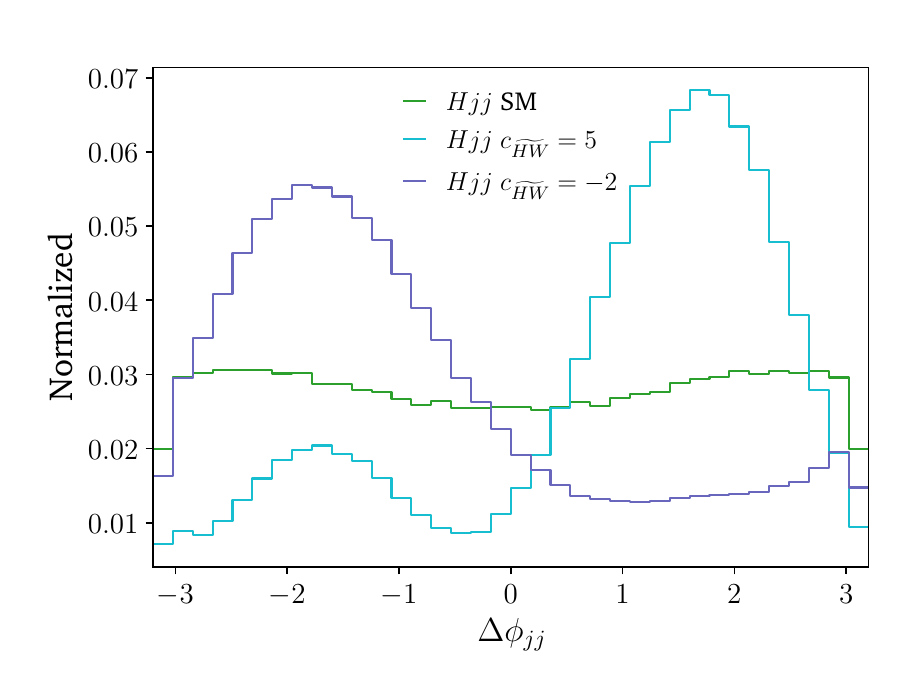}
    \caption{Profile of $-\ln L(c_{\widetilde{HW}})$ for the SM, CP-odd($+$), and CP-odd($-$) pseudo-data. We train the TraCFM unfolding network on the SM or conditionally on all five hypotheses and compare the results to the reco-level fit. For the SM, we also show the fit to the unfolded background-subtracted data (orange). Bottom right: morphing templates in $\Delta\phi_{jj}$.}
    \label{fig:nll_all}
\end{figure}

For the fit on unfolded data, we unfold each of the three signal-only datasets with the TraCFM network trained on SM events, and with the network trained conditional on all five hypotheses but using the SM label throughout. After unfolding, we compute $\Delta\phi_{qq}$ and fill it into four bins of equal width in $[-\pi,\pi)$, matching the corresponding ATLAS analysis~\cite{ATLAS:2025hki}. We then perform a maximum likelihood fit, where we parameterize the expected distributions of $\Delta\phi_{qq}$ as a function of $c_{\widetilde{HW}}$, constructed using the three templates of parton-level simulations. This gives us the maximum likelihood estimate $\hat{c}_{\widetilde{HW}}$ shown in Tab.~\ref{tab:MLEvalues}.

We compare this outcome to a fit to the reco-level pseudo-data. In this case, we fill $\Delta\phi_{jj}$ into the same four bins and fit them with the same procedure as above, where we parameterize the expected distributions of $\Delta\phi_{jj}$ as a function of $c_{\widetilde{HW}}$, now using the three templates at reco level, and obtain again the best fit value for the Wilson coefficient shown in the last row of Tab.~\ref{tab:MLEvalues}. The best-fit values and the statistical uncertainties agree well between the unfolded parton-level and reco-level inference. The largest difference to the reco-level result, $0.06$ for CP-odd($-$) with the SM-trained network, is below 3\% of the width of the 68\% CL interval. The unfolded intervals tend to be narrower than at reco level, reflecting the more informative parton-level $\Delta\phi_{qq}$ distribution.
  
Finally, we repeat the SM fit on the background-subtracted data of Sec.~\ref{sec:back}, \ie SM signal together with the $t\bar{t}$ and $WWjj$ backgrounds, unfolded with the SM-trained network and every event weighted by $1-w(\xr)$. We find $\hat{c}_{\widetilde{HW}} = 0.025$, compared with $0.006$ for the signal-only unfolding, a shift of less than 1\% of the width of the 68\% CL interval. The background removal therefore does not bias the measurement at this level.
  
We show the corresponding likelihood profiles in Fig.~\ref{fig:nll_all}. For the CP-odd pseudo-data, they show a second local minimum at the opposite sign of $c_{\widetilde{HW}}$. The interference of the CP-odd operator with the SM is antisymmetric in $\Delta\phi_{jj}$ and does not contribute to the total rate. It is only the asymmetry between the positive and negative $\Delta\phi_{jj}$ that distinguishes the two signs, lifting the minimum at the wrong sign.

\section{Outlook}
\label{sec:conclusions}

Weak boson fusion Higgs production allows us to test the Standard Model through the tagging jet kinematics and the central electroweak scattering, making an exciting case for unfolding. It provides three distinct challenges: First, for Higgs decays through $W$ bosons, unfolding the missing momentum and the transverse mass is a technical challenge. Second, when searching for effects from dimension-6 operators, with and without CP violation, it is critical to control the bias from the training data. Finally, the top-pair and continuum $WWjj$ backgrounds make up more than $80\%$ of the fiducial events, so we have to combine unfolding with a statistical background removal. 

We tackled these three challenges with generative unfolding, technically a conditional flow matching network including a transformer, TraCFM. A suitable phase space parameterization is essential for precise unfolding. Unfolding the transverse mass directly, together with an ellipse construction for the quark transverse momenta, performs well over the entire phase space. This construction guarantees transverse momentum conservation and the tagging jet selection for every unfolded event. Unfolding the lepton features as shifts with respect to their reco-level values reduced that task to learning the narrow detector response.

Second, we have quantified the dependence of the unfolded distributions on the simulation used for training. Training conditionally on the SM and the four SMEFT samples combined lead to closure at the percent level. Since the detector response is shared between all theory hypotheses, the conditionally trained network benefits from the combined training statistics. In reality, when the label is unknown, the envelope over all labels provides a systematic uncertainty. It stayed below the statistical uncertainty for the Higgs decay observables and reached three times the statistical uncertainty for the tagging jets, where the reco-level event constrains the parton-level kinematics the least.

Finally, we have removed the large backgrounds using a classifier-based reweighting. Although the signal amounts to only 14\% of the data, the background-reweighted unfolded distributions agree with the parton-level truth at the percent level in the bulk. We tested our unfolding by extracting the CP-odd Wilson coefficient from the signed azimuthal angle of the unfolded tagging quarks and from the signed azimuthal angle of the reconstructed tagging jets. The best-fit values and uncertainty bands agree at the $3\%$ level. Extracting the coefficient from the unfolded background-subtracted SM data shifts the best-fit value by only about 1\% of the 68\% CL interval, so the background removal does not bias the measurement.
  
\subsection*{Acknowledgments}
The Heidelberg group is supported by the KISS consortium (05D23GU4) funded by BMFTR in the ErUM-Data action plan and the Deutsche Forschungsgemeinschaft (DFG, German Research Foundation) under grant 396021762 -- TRR~257 \textsl{Particle Physics Phenomenology after the Higgs Discovery}. JMV is particularly grateful to TRR~257 for the extended funding, so this paper could be finished. AB and BM gratefully acknowledge the continuous support from LPNHE, CNRS/IN2P3, Sorbonne Université and Université de Paris Cité. SPS acknowledges support from the Alexander von Humboldt foundation as a Feodor Lynen Fellow and from the DOE grant DE-SC0010008. 

\subsection*{Code availability}
The code for this paper can be found at \url{https://github.com/heidelberg-hepml/WBFUnfolding}.

\clearpage
\appendix

\section{Hyperparameters}
\label{sec:hyperparams}

\begin{table}[hp!]
\centering
\begin{footnotesize}
\begin{tabular}{l|cc}
\toprule
 & SM training & SM+EFTs training \\
\midrule
Network & \multicolumn{2}{c}{TraCFM} \\
Embedding dimension & \multicolumn{2}{c}{80} \\
Attention heads & \multicolumn{2}{c}{4} \\
Encoder / decoder layers & \multicolumn{2}{c}{8 / 8} \\
Feed-forward dimension & \multicolumn{2}{c}{320} \\
Dropout & \multicolumn{2}{c}{0} \\
Label embedding dimension & -- & 8 \\
Trainable parameters & 1\,454\,162 & 1\,454\,202 \\
\midrule
Train / test / val split & \multicolumn{2}{c}{75\% / 20\% / 5\%} \\
Training events & 1.47M SM & 7.55M [SM, CP-even($\pm$), CP-odd($\pm$)] balanced\\
Optimizer & \multicolumn{2}{c}{Adam} \\
Learning rate & \multicolumn{2}{c}{$10^{-4}$} \\
LR schedule & \multicolumn{2}{c}{ReduceLROnPlateau, factor 0.5, patience 25 epochs} \\
Batch size & \multicolumn{2}{c}{4096} \\
Early stopping & \multicolumn{2}{c}{patience 75 epochs, capped at $10^6$ iterations} \\
Training iterations & 185\,760 & 777\,746 \\
Network used & \multicolumn{2}{c}{final (at early stopping)} \\
ODE solver & \multicolumn{2}{c}{adaptive dopri5, rtol $10^{-5}$, atol $10^{-7}$} \\
\bottomrule
\end{tabular}
\end{footnotesize}
\caption{Hyperparameters of the TraCFM networks, trained on the SM signal (Secs.~\ref{sec:wbf_sig} and~\ref{sec:back}) and on the SM and the four EFT samples with label conditioning (Secs.~\ref{sec:wbf_eft} and~\ref{sec:wbf_env}).}
\label{tab:hyper_unf}

\vspace{1.5em}

\begin{footnotesize}
\begin{tabular}{l|cc}
\toprule
 & $t\bar{t}$ reweighting & background subtraction \\
\midrule
Network & \multicolumn{2}{c}{Transformer encoder} \\
Inputs & 24 reco-level features (incl.\ $n_j$) & 23 reco-level features \\
Preprocessing & \multicolumn{2}{c}{Gaussianized} \\
Embedding dimension & \multicolumn{2}{c}{128} \\
Attention heads & \multicolumn{2}{c}{8} \\
Encoder layers & \multicolumn{2}{c}{6} \\
Feed-forward dimension & \multicolumn{2}{c}{128} \\
Dropout & 0 & 0.1 \\
Trainable parameters & \multicolumn{2}{c}{$\approx 598$k} \\
\midrule
Classes & particle-level vs.\ parton-level cuts & data vs.\ MC background \\
Train / val events & 0.86M / 0.10M & 2.95M / 0.17M \\
Optimizer & \multicolumn{2}{c}{Adam} \\
Weight decay & 0 & $10^{-4}$ \\
Learning rate & \multicolumn{2}{c}{$10^{-4}$} \\
LR schedule & \multicolumn{2}{c}{ReduceLROnPlateau, factor 0.5, patience 10 epochs} \\
Batch size & \multicolumn{2}{c}{1024} \\
Training iterations & $5\times10^4$ & $10^5$ \\
\midrule
Ensemble & 5 initializations & 5 initializations $\times$ 3 folds \\
Normalization & mean weight 1 & background yield per member \\
\bottomrule
\end{tabular}
\end{footnotesize}
\caption{Hyperparameters of the classifiers for the $t\bar{t}$ reweighting from particle-level to parton-level cuts and for the background subtraction, Sec.~\ref{sec:back}.}
\label{tab:hyper_clf}
\end{table}

\clearpage
\section{Transverse mass parameterization}
\label{sec:appendix_mt_param}

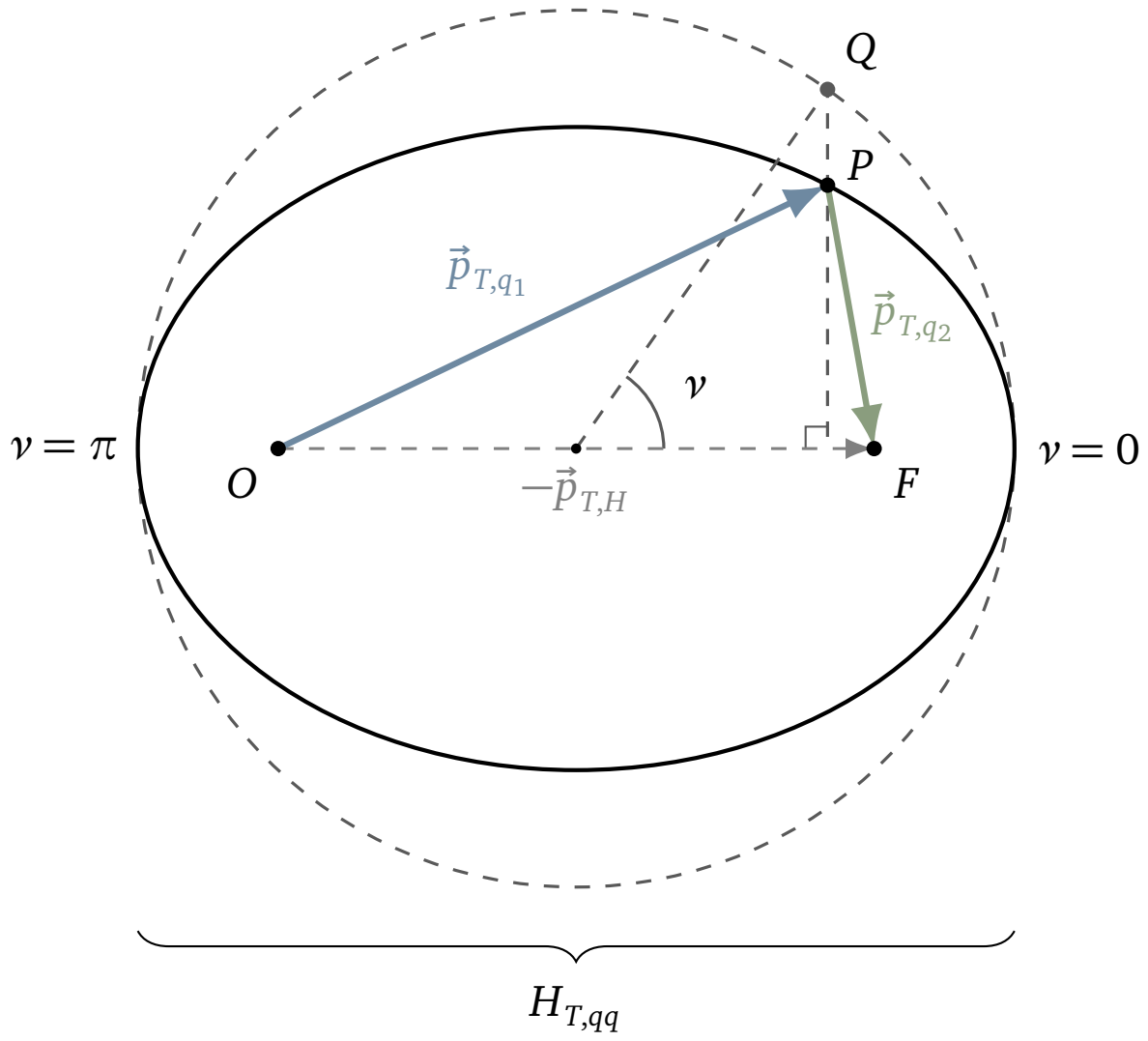
\begin{figure}[b!]
\centering
\input{tikz/transverse_mass_param_ellipse.tex}
\caption{Quark pairs in the transverse plane for fixed $H_{T,qq}$. The quarks recoil against the Higgs boson, so the tip $P$ of $\vec p_{T,q_1}$ lies on the ellipse with foci $O$ and $F$ and major axis $H_{T,qq}$. The dashed circle is the auxiliary circle of the ellipse, centred at $-\vec p_{T,H}/2$ with radius $H_{T,qq}/2$. The eccentric anomaly $\nu$ is the angle of the point $Q$ on this circle, and $P$ is $Q$ moved perpendicular to the major axis onto the ellipse.}
\label{fig:ellipse}
\end{figure}

We complete here the construction of parton-level events from the features of Eq.\eqref{eq:transverse_mass_param}. Of these, the eight lepton and missing-momentum features
\begin{align}
    \Big\{ \;
    p_{T,\ell_1},\;
    \phi_{\ell_1},\;
    \eta_{\ell_1},\;
    \Delta\phi_{\ell\ell},\;
    \Delta\eta_{\ell\ell},\;
    m_T,\;
    r_{\ell\ell},\;
    \Delta\phi_{\ell\ell,\text{miss}}
    \Big\}
\end{align}
are unfolded as shifts with respect to their reco-level values, Eq.\eqref{eq:delta_lepton}, while the four quark features $\{\Delta\eta_{qq},\, \bar\eta_q,\, H_{T,qq},\, \nu\}$ are unfolded directly. The allowed ranges of the quark features depend on $p_{T,H}$ and on each other, as shown below. At reco level, momentum balance holds only approximately, and the reco-level values can lie outside these ranges. For instance, $p_{T,H} > H_{T,jj}$ is possible, in which case the ellipse defined below does not exist. A shift with respect to the reco-level value is then not defined.

The dilepton mass follows as $m_{\ell\ell} = r_{\ell\ell}\, m_T$, and with it $p_{T,\ell_2}$ from Eq.\eqref{eq:ptl2_mt}, $\phi_{\ell_2} = \phi_{\ell_1} - \Delta\phi_{\ell\ell}$, and $\eta_{\ell_2} = \eta_{\ell_1} - \Delta\eta_{\ell\ell}$. If $p_{T,\ell_2} > p_{T,\ell_1}$, the two leptons are relabeled, which flips the signs of $\Delta\phi_{\ell\ell}$ and $\Delta\eta_{\ell\ell}$. Solving Eq.\eqref{eq:transverse_mass} with $m_\text{miss}=0$ for the missing transverse energy gives
\begin{align}
    E_{T,\text{miss}} = \frac{m^2_{T} - m^2_{\ell\ell}}{2\left(E_{T,\ell\ell} - p_{T,\ell\ell}\cos\Delta\phi_{\ell\ell,\text{miss}}\right)}
    \qquad \text{and} \qquad
    \phi_\text{miss} = \phi_{\ell\ell} - \Delta\phi_{\ell\ell,\text{miss}}\; ,
\end{align}
which is positive for every $r_{\ell\ell} < 1$, since $E_{T,\ell\ell} > p_{T,\ell\ell}$. The Higgs transverse momentum is then $\vec p_{T,H} = \vec p_{T,\ell\ell} + \vec p_{T,\text{miss}}$.

At parton level, the quarks recoil exactly against the Higgs boson, Eq.\eqref{eq:higgs_propagator}. Squaring this relation gives
\begin{align}
    p_{T,q_1}\, p_{T,q_2} = \frac{H^2_{T,qq} - p^2_{T,H}}{2\left(1-\cos\Delta\phi_{qq}\right)}\; ,
\end{align}
so for a given $\vec p_{T,H}$ the quark transverse momenta have two degrees of freedom. We describe them by the scalar sum $H_{T,qq}$ and an angle $\nu$, which encodes both $\Delta\phi_{qq}$ and which quark is harder. Fixing $H_{T,qq}$, the tip of $\vec p_{T,q_1}$ lies on an ellipse with foci at the endpoints of $-\vec p_{T,H}$ and major axis $H_{T,qq}$ (see Fig.~\ref{fig:ellipse}), and each point on the ellipse corresponds to a valid quark pair. For $p_{T,H} = H_{T,qq}$ the ellipse degenerates to a segment and the quarks are collinear in the transverse plane. For $p_{T,H} = 0$ it degenerates to a circle and the quarks have equal transverse momenta. We label the points on the ellipse by the eccentric anomaly $\nu \in (-\pi, \pi]$. We use the eccentric anomaly rather than the polar angle of $P$ around a focus because of the collinear limit. When the ellipse degenerates to a segment, the polar angle no longer distinguishes quark pairs with different transverse momenta. The eccentric anomaly is defined on the auxiliary circle, which does not degenerate, so it labels all quark pairs for every value of $p_{T,H}$. The distance of $P$ to either focus is linear in $\cos\nu$, which gives
\begin{align}
  p_{T,q_{1,2}} = \tfrac12\left(H_{T,qq} \pm p_{T,H}\cos\nu\right)
  \qquad \text{and} \qquad
  \Delta\phi_{qq} = 2\operatorname{atan2}\!\left(\sqrt{H_{T,qq}^2-p_{T,H}^2},\; p_{T,H}\sin\nu\right) \; ,
  \label{eq:loop}
\end{align}
with $\Delta\phi_{qq}$ mapped into $(-\pi,\pi]$. The quark azimuthal angles follow from the recoil condition,
\begin{align}
    \phi_{q_2} = \phi_H + \pi - \operatorname{atan2}\left(p_{T,q_1}\sin\Delta\phi_{qq},\, p_{T,q_2} + p_{T,q_1}\cos\Delta\phi_{qq}\right)
    \qquad \text{and} \qquad
    \phi_{q_1} = \phi_{q_2} + \Delta\phi_{qq} \; .
\end{align}
All simulated events pass the selection cuts, while unfolded events can populate regions outside them. A classifier separating unfolded from simulated events can then rely on these regions alone and lose sensitivity to the remaining kinematic features. We therefore enforce the quark selection by construction. Both quarks pass the threshold $p_{T,\text{cut}} = 30$~GeV if and only if the softer one does,
\begin{align}
  H_{T,qq} - p_{T,H}\,|\cos\nu| > 2p_{T,\text{cut}}\; .
\end{align}
This translates into a lower bound on $H_{T,qq}$ and a constraint on $\nu$,
\begin{align}
  H_{T,qq} > H_0 \equiv \max\left(p_{T,H},\, 2p_{T,\text{cut}}\right)
  \qquad \text{and} \qquad
  |\cos\nu| < \cos\nu_0 \equiv \min\left(1,\; \frac{H_{T,qq} - 2p_{T,\text{cut}}}{p_{T,H}}\right) \;.
\end{align}
Geometrically, the constraint on $\nu$ removes the two ends of the major axis of the ellipse, where one of the quarks is softest. We map the allowed range $\nu_0 < |\nu| < \pi - \nu_0$ onto the full circle,
\begin{align}
  u = \pi\,\frac{\nu - \sign(\nu)\,\nu_0}{\pi - 2\nu_0} \in (-\pi,\pi) \;,
\end{align}
and unfold $u$ instead of $\nu$. Once the quark transverse momenta and $\Delta\phi_{qq}$ are fixed, the invariant mass of the massless quark pair,
\begin{align}
  m_{qq}^2 = 2\,p_{T,q_1}p_{T,q_2}\left(\cosh\Delta\eta_{qq} - \cos\Delta\phi_{qq}\right) \;,
\end{align}
grows monotonically with $|\Delta\eta_{qq}|$. The cuts on $m_{qq}$ and on the rapidity gap therefore combine into one lower bound,
\begin{align}
  |\Delta\eta_{qq}| > \Delta\eta_\text{min} \equiv \max\left(2.1,\;
  \operatorname{arccosh}\left( \max\left[1,\; \cos\Delta\phi_{qq} + \frac{m_{qq,\text{cut}}^2}{2\,p_{T,q_1}p_{T,q_2}}\right] \right)\right) \;,
  \label{eq:deta_min}
\end{align}
with $m_{qq,\text{cut}} = 450$~GeV. The inner maximum covers the case where the mass cut is already satisfied for $\Delta\eta_{qq} = 0$. Finally, $|\eta_{q_{1,2}}| < 4.5$ holds if and only if $|\bar\eta_q| < 4.5 - |\Delta\eta_{qq}|/2$. This range is never empty, because $p_{T,q_{1,2}} > p_{T,\text{cut}}$ implies
\begin{align}
  \Delta\eta_\text{min} \le \operatorname{arccosh}\left(1 + \frac{m_{qq,\text{cut}}^2}{2\,p_{T,\text{cut}}^2}\right) \approx 5.4 < 9 \; .
\end{align}
The lepton maps enforce $p_{T,\ell_1} > 22$~GeV, through $\log(p_{T,\ell_1} - 22~\text{GeV})$, $|\Delta\phi_{\ell\ell}| < 1.4$, through $\operatorname{artanh}(\Delta\phi_{\ell\ell}/1.4)$, and $m_{\ell\ell} < m_T$, through the logit of $r_{\ell\ell}$. The parameterization does not enforce $p_{T,\ell_2} > 15$~GeV, $|\eta_\ell| < 2.5$, $m_{\ell\ell} > 10$~GeV, $\Delta R(\ell,\ell) > 0.1$, or the lepton-jet separation $\Delta R(\ell,j) > 0.4$.

Since the bounds of later features depend on earlier ones, an unfolded event is completed in the following order:
\begin{enumerate}
    \item Leptons: $p_{T,\ell_1}$, $\phi_{\ell_1}$, $\eta_{\ell_1}$, $\Delta\phi_{\ell\ell}$, $\Delta\eta_{\ell\ell}$, $m_T$, and $r_{\ell\ell}$ are obtained from the unfolded shifts, and the second lepton follows as described above.
    \item Missing transverse momentum: $E_{T,\text{miss}}$ and $\phi_\text{miss}$ follow from $m_T$ and $\Delta\phi_{\ell\ell,\text{miss}}$.
    \item Higgs recoil: $\vec p_{T,H} = \vec p_{T,\ell\ell} + \vec p_{T,\text{miss}}$.
    \item Quark transverse momenta: $H_{T,qq}$ above $H_0$, then $\nu$ from $u$. Eq.\eqref{eq:loop} gives $p_{T,q_1}$, $p_{T,q_2}$, and $\Delta\phi_{qq}$.
    \item Quark azimuthal angles: $\phi_{q_2}$ from the recoil against $\vec p_{T,H}$, and $\phi_{q_1} = \phi_{q_2} + \Delta\phi_{qq}$.
    \item Quark pseudorapidities: $|\Delta\eta_{qq}|$ above $\Delta\eta_\text{min}$, then $\bar\eta_q$ within $|\bar\eta_q| < 4.5 - |\Delta\eta_{qq}|/2$, and $\eta_{q_{1,2}} = \bar\eta_q \pm \Delta\eta_{qq}/2$.
\end{enumerate}

\section{Unfolding envelope contributions}
\label{sec:wbf_env_contributions}

\begin{figure}[h]
    \centering
    \includegraphics[width=0.445\textwidth, page=1]{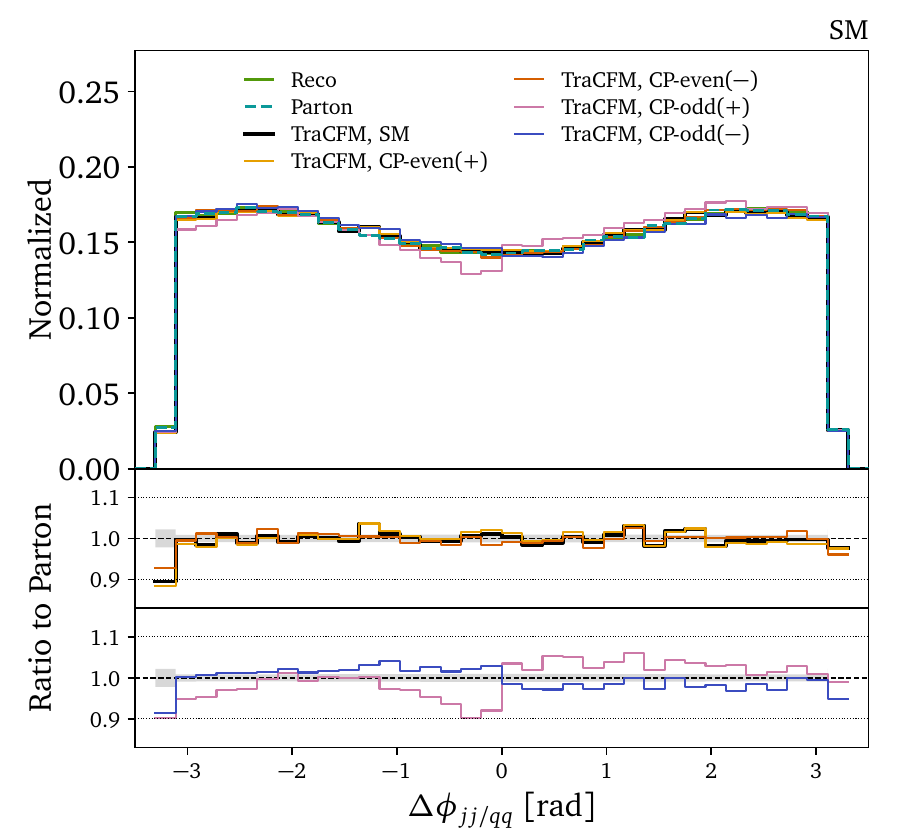}
    \includegraphics[width=0.445\textwidth, page=5]{figs/allsamples/label_contributions_sm_cpevenpos_cpoddpos.pdf}
    \caption{Contributions to the envelope of Fig.~\ref{fig:trainSMEvenOdd-unfoldSMEvenOdd_spread} for $\Delta\phi_{jj}$ for the SM (left), and CP-odd($+$) samples unfolded with each of the five labels (with the correct label drawn thicker). The ratio panels show each unfolding divided by the parton truth, whose statistical uncertainty is shown as a gray band.}
    \label{fig:label_contributions_sm_cpevenpos_cpoddpos}
\end{figure}

\clearpage
\bibliography{tilman,literature,literature2}
\end{document}

%% file: incl_settings.tex
\usepackage[utf8]{inputenc} 
\usepackage[T1]{fontenc} 	
\usepackage[english]{babel} 

\usepackage[bitstream-charter]{mathdesign}
\usepackage{geometry} 		
\usepackage{amsmath} 		
\usepackage{mathtools} 		
\usepackage{float} 			
\usepackage{graphicx} 		
\usepackage{tabularx} 		
\usepackage{booktabs} 		
\usepackage{color, xcolor} 	
\usepackage{pdfpages} 		
\usepackage{extarrows} 		
\usepackage{multirow} 		
\usepackage{multicol} 		
\usepackage{caption} 		
\usepackage{subcaption} 	
\usepackage{enumitem} 		
\usepackage{xspace} 		
\usepackage{stackrel} 		
\usepackage{tikz}
\usetikzlibrary{calc,decorations.pathreplacing,arrows.meta}
\tikzset{arrow/.style={-{Stealth[length=2mm]}}}
\definecolor{Bcolor}{HTML}{0B9999}
\definecolor{Gcolor}{HTML}{990B0B}
\usepackage{braket} 		
\usepackage{bm} 			
\usepackage{tensor} 		
\usepackage{slashed} 		
\usepackage{siunitx} 
\usepackage{lastpage} 		
\usepackage{cite} 			
\usepackage[normalem]{ulem} 
\usepackage{fontawesome} 	
\usepackage{tocloft} 		
\usepackage{titlesec} 		
\usepackage{doi} 			
\usepackage[most]{tcolorbox} 					
\usepackage[capitalize]{cleveref} 	
\usepackage[nottoc, notlot, notlof]{tocbibind} 	
\usepackage[ruled, vlined]{algorithm2e} 		
\usepackage{makecell}
\usepackage{pifont}
\usepackage{stackrel}

\hypersetup{
	pdftitle={Unfolding Weak Boson Fusion},
	pdfauthor={Us},
    linktoc=section,
	breaklinks=true,			
	colorlinks=true, 			
	linkcolor={red!50!black}, 	
	citecolor={blue!50!black}, 	
	urlcolor={blue!80!black} 	
} 

\DeclareSymbolFont{usualmathcal}{OMS}{cmsy}{m}{n}
\DeclareSymbolFontAlphabet{\mathcal}{usualmathcal}

\SetArgSty{textnormal}
\SetKwComment{Comment}{{\small\#}~}{}
\SetCommentSty{mycommfont}

\setitemize{itemsep=2pt, parsep=0pt} 				
\setenumerate{itemsep=2pt, parsep=0pt} 				
\crefname{section}{Sec.}{Secs.}
\crefname{equation}{Eq.}{Eqs.}
\crefname{figure}{Fig.}{Figs.}
\crefname{table}{Tab.}{Tabs.}
\Crefname{section}{Section}{Sections}
\Crefname{equation}{Equation}{Equations}
\Crefname{figure}{Figure}{Figures}
\Crefname{table}{Table}{Tables}

\usetikzlibrary{arrows, shapes.geometric, arrows.meta, shapes, decorations.pathreplacing, fit, patterns, patterns.meta}

\definecolor{Rcolor}{HTML}{E99595}
\definecolor{Gcolor}{HTML}{C5E0B4}
\definecolor{Bcolor}{HTML}{9DC3E6}
\definecolor{Ycolor}{HTML}{FFE699}

\tikzstyle{expr} = [circle, minimum width=1.8cm, minimum height=1.8cm, text centered, align=center, inner sep=0, draw,font=\LARGE]
\tikzstyle{txt_huge} = [align=center, font=\Huge, scale=2]
\tikzstyle{txt} = [align=center, font=\LARGE]
\tikzstyle{cinn} = [double arrow, double arrow head extend=0cm, double arrow tip angle=130, shape border rotate=90, inner sep=0, align=center, minimum width=2.1cm, minimum height=2.3cm, fill=Bcolor, draw,font=\LARGE]
\tikzstyle{cinn_black} = [cinn, minimum height=2.5cm, fill=black]
\tikzstyle{arrow} = [thick,-{Latex[scale=1.0]}, line width=0.2mm, color=black]
\tikzstyle{line} = [thick, line width=0.2mm, color=black]
\tikzstyle{loss} = [rectangle, align=center,  minimum width=1.8cm, minimum height=1.5cm,fill=Rcolor,font=\LARGE, rounded corners]
\tikzstyle{xt} = [rectangle, align=center,  minimum width=4cm, minimum height=1.5cm,fill=Gcolor,font=\Large, rounded corners]
\tikzstyle{xts} = [rectangle, align=center,  minimum width=1cm, minimum height=1.5cm,fill=Gcolor,font=\Large, rounded corners]

%% file: incl_shortcuts.tex
\newcommand{\ie}{i.e.\@\xspace} 

\newcommand{\qqquad}{\qquad\quad}
\newcommand{\qqqquad}{\qquad\qquad}

\newcommand{\sign}{\operatorname{sign}} 

\newcommand{\normal}{\mathcal{N}}

\newcommand{\ope}{\mathcal{O}}

\newcommand{\loss}{\mathcal{L}} 

\newcommand{\psim}{p_\text{sim}}
\newcommand{\ps}{p_\text{signal}}
\newcommand{\pbg}{p_\text{bkg}}

\newcommand{\arXiv}[2][]{%
\ifthenelse{\equal{#1}{}}%
{\href{http://arxiv.org/abs/#2}{arXiv:#2}}%
{\href{http://arxiv.org/abs/#2}{arXiv:#2~[#1]}}}

\usepackage{xspace}

\newcommand{\gev}{\ensuremath{\,\text{GeV}}\xspace}
\newcommand{\tev}{\ensuremath{\,\text{TeV}}\xspace}

\def\slashchar#1{\setbox0=\hbox{$#1$} 
\dimen0=\wd0 
\setbox1=\hbox{/} \dimen1=\wd1 
\ifdim\dimen0>\dimen1 
\rlap{\hbox to \dimen0{\hfil/\hfil}} 
#1 
\else 
\rlap{\hbox to \dimen1{\hfil$#1$\hfil}} 
/ 
\fi}

\renewcommand{\d}{{\rm d}}

\newcommand{\tikznode}[2]{%
\ifmmode%
\tikz[remember picture,baseline=(#1.base),inner sep=0pt]{\node (#1) {$#2$};}%
\else \tikz[remember picture,baseline=(#1.base),inner sep=0pt]{\node (#1) {#2};}%
\fi}

\def\mathswitchr#1{\relax\ifmmode{\text{#1}}\else$\text{#1}$\xspace\fi} \def\mathswitch#1{\relax\ifmmode#1\else$#1$\xspace\fi}

\newcommand{\ai}{\alpha_i}

\newcommand{\pd}{p_\text{data}}

\newcommand{\pmd}{p_\text{model}}
\newcommand{\xr}{x_\text{reco}}
\newcommand{\xp}{x_\text{part}}

\newcommand{\mwith}{\text{with}}

%% file: tikz/transverse_mass_param_ellipse.tex
\begin{tikzpicture}[font=\footnotesize]
\pgfmathsetmacro{\sa}{3}
\pgfmathsetmacro{\sb}{2.2}
\pgfmathsetmacro{\fc}{sqrt(\sa*\sa-\sb*\sb)}
\pgfmathsetmacro{\nuang}{55}

\coordinate (C) at (0,0);
\coordinate (O) at (-\fc,0);
\coordinate (F) at (\fc,0);
\coordinate (P) at ({\sa*cos(\nuang)},{\sb*sin(\nuang)});
\coordinate (Q) at ({\sa*cos(\nuang)},{\sa*sin(\nuang)});

\draw[black!65, dashed, semithick] (C) circle (\sa);
\draw[thick] (C) ellipse ({\sa} and {\sb});
\coordinate (N) at ({\sa*cos(\nuang)},0);
\draw[black!65, dashed, semithick] (C) -- (Q) -- (N);
\draw[black!65] ($(N)+(-0.15,0)$) -- ($(N)+(-0.15,0.15)$) -- ($(N)+(0,0.15)$);
\draw[black!65, semithick] ($(C)+(0.6,0)$) arc (0:\nuang:0.6);
\node at ($(C)+({\nuang/2}:0.9)$) {$\nu$};

\draw[arrow, dashed, gray] (O) -- (F) node[midway, below] {$-\vec p_{T,H}$};
\draw[arrow, very thick, Bcolor!70!black] (O) -- (P) node[midway, above left] {$\vec p_{T,q_1}$};
\draw[arrow, very thick, Gcolor!70!black] (P) -- (F) node[midway, right] {$\vec p_{T,q_2}$};

\fill (C) circle (1pt);
\fill (O) circle (1.5pt) node[below left] {$O$};
\fill (F) circle (1.5pt) node[below right] {$F$};
\fill (P) circle (1.5pt) node[right, yshift=4pt] {$P$};
\fill[black!65] (Q) circle (1.5pt) node[above right, black] {$Q$};

\node[right] at (\sa,0) {$\nu=0$};
\node[left] at (-\sa,0) {$\nu=\pi$};

\draw[decorate, decoration={brace, amplitude=6pt, mirror}]
  (-\sa,{-\sa-0.3}) -- (\sa,{-\sa-0.3})
  node[midway, below=7pt] {$H_{T,qq}$};
\end{tikzpicture}